\documentclass{article}

\usepackage{amsmath, amsthm, amssymb}
\usepackage{graphicx}
\usepackage{verbatim}
\usepackage{natbib}
\usepackage{caption}
\usepackage{subcaption}
\usepackage{fancyvrb}
\usepackage[inline]{enumitem}
\usepackage{relsize}
\usepackage{hyperref}
\usepackage[margin=1.5in]{geometry}
\hypersetup{colorlinks,citecolor=blue,urlcolor=blue,linkcolor=blue}
\usepackage{diagbox}
\usepackage{stefan_tex}
\usepackage{float}
\usepackage{footmisc}

\usepackage{multirow}

\usepackage[utf8]{inputenc}
\usepackage[english]{babel}

\usepackage{accents}
\newcommand{\ubar}[1]{\underaccent{\bar}{#1}}

\usepackage{pgfplots}
\usepackage{tikzscale}

\pgfplotsset{compat=newest}
\usepgfplotslibrary{groupplots}
\usepgfplotslibrary{polar}
\usepgfplotslibrary{smithchart}
\usepgfplotslibrary{statistics}
\usepgfplotslibrary{dateplot}
\usepgfplotslibrary{external}
\usetikzlibrary{arrows.meta}
\usetikzlibrary{backgrounds}
\usepgfplotslibrary{patchplots}
\usepgfplotslibrary{fillbetween}
\pgfplotsset{%
layers/standard/.define layer set={%
    background,axis background,axis grid,axis ticks,axis lines,axis tick labels,pre main,main,axis descriptions,axis foreground%
}{grid style= {/pgfplots/on layer=axis grid},%
    tick style= {/pgfplots/on layer=axis ticks},%
    axis line style= {/pgfplots/on layer=axis lines},%
    label style= {/pgfplots/on layer=axis descriptions},%
    legend style= {/pgfplots/on layer=axis descriptions},%
    title style= {/pgfplots/on layer=axis descriptions},%
    colorbar style= {/pgfplots/on layer=axis descriptions},%
    ticklabel style= {/pgfplots/on layer=axis tick labels},%
    axis background@ style={/pgfplots/on layer=axis background},%
    3d box foreground style={/pgfplots/on layer=axis foreground},%
    },
}

\usepackage{tikz}
\usetikzlibrary{shapes,decorations,arrows,fit,positioning, calc}
\tikzset{
    state/.style ={ellipse, draw, minimum width = 0.7 cm},
    el/.style = {inner sep=2pt, align=left, sloped}
}

\usepackage{algorithm2e}
\RestyleAlgo{boxruled}
\LinesNumbered
\DontPrintSemicolon
\SetAlFnt{\normalsize}
\SetKwInOut{Input}{Input}

\usepackage{soul}

\theoremstyle{plain}
\newtheorem{proposition}{Proposition}
\newtheorem{corollary}{Corollary}
\newtheorem{lemma}{Lemma}
\newtheorem{theorem}{Theorem}

\theoremstyle{definition}
\newtheorem{example}{Example}
\newtheorem{assumption}{Assumption}

\theoremstyle{remark}

\author{
\makebox[45mm]{Dean Eckles} \\ MIT, USA \and
\makebox[45mm]{Nikolaos Ignatiadis\thanks{Authors
are listed in alphabetical order. Corresponding author: \url{ignat@uchicago.edu}}} \\ University of Chicago, USA \and
\makebox[45mm]{Stefan Wager} \\  Stanford University, USA \and
\makebox[45mm]{Han Wu} \\ Two Sigma Investments, USA}

\date{\ifcase\month\or
January\or February\or March\or April\or May\or June\or
July\or August\or September\or October\or November\or December\fi \ \number%
\year\ \  }

\title{Noise-Induced Randomization in \\ Regression Discontinuity Designs\thanks{
We thank
Alex D'Amour,
Timothy Armstrong,
Matias Cattaneo,
Jan Gleixner,
Jiaying Gu,
David Hirshberg,
Guido Imbens,
Roger Koenker,
Michal Koles\'ar,
Fan Li,
Charles Manski,
Fabrizia Mealli,
Jamie Robins,
Jasjeet Sekhon,
Johan Ugander,
and Jos\'e Zubizarreta
for helpful suggestions and discussions. 
We are grateful to the editor, Prof. Paul Fearnhead, the associate editor, and reviewers for their thorough and constructive feedback, which substantially improved this manuscript.
Some of the computing for this project was performed 
with support from the Research Computing Centers at the University of Chicago and Stanford University. We acknowledge support from National Science Foundation grant DMS-1916163.
D.E. was partially supported by funding from the MIT--IBM Watson AI Lab.
}}

\begin{document}

\maketitle

\begin{abstract}
Regression discontinuity designs assess causal effects in settings where treatment is determined by whether an observed running variable crosses a pre-specified threshold. Here we propose a new approach to identification, estimation, and inference in regression discontinuity designs that uses knowledge about exogenous noise (e.g., measurement error) in the running variable. In our strategy, we weight treated and control units to balance a latent variable of which the running variable is a noisy measure. Our approach is driven by effective randomization provided by the noise in the running variable, and complements standard formal analyses that appeal to continuity arguments while ignoring the stochastic nature of the assignment mechanism.\\

\end{abstract}

\section{Introduction}

Regression discontinuity designs rely on known,
discontinuous treatment assignment mechanisms to identify causal effects
\citep*{hahn2001identification,imbens2008regression,thistlethwaite1960regression}:
There is a running variable $Z_i \in \RR$ such that unit $i$
gets assigned treatment $W_i \in \cb{0, \, 1}$ whenever the running variable exceeds a cutoff
$c \in \RR$, i.e., $W_i = 1(Z_i \geq c)$, and we estimate treatment effects by
comparing units with $Z_i$ just above or below $c$.
For example, in an educational setting where admission to a program hinges on a test score
exceeding some cutoff, we could evaluate the effect of the program on marginal admits by
comparing outcomes for students whose test scores fell right above and below the cutoff.
Over the past decades, regression discontinuity designs have become one of the most
widely used methods for causal inference, especially in the social sciences \citep*{currie2020technology}.

Explanations and qualitative justifications of identification in regression discontinuity designs often appeal to implicit, local randomization: Many factors outside the control of decision-makers determine the running variable $Z_i$ such that if some unit barely clears the eligibility cutoff for the intervention then the same unit could also plausibly have failed to clear the cutoff with a different realization of these chance factors \citep{lee2010regression}.
This is sometimes illustrated by reference to sampling error or other errors in measurement that cause units to have a measured running variable just above or just below the threshold. In our educational setting, there may be a group of marginal students who might barely pass or fail the test due to unpredictable variation in their test score, resulting in an effectively exogenous treatment assignment rule. Likewise, medical assays frequently involve a degree of random measurement error, whether because of sampling techniques or other sources of random variation \citep{bor2014regression}.

Most formal and practical approaches to identification, estimation, and inference for treatment effects in regression discontinuity designs,
however, do not use exogenous noise in the running variable to drive inference. Instead, following
\citet*{hahn2001identification}, the dominant approach relies on a continuity argument.
As in \citet{imbens2008regression}, we assume potential outcomes $\cb{Y_i(0), \, Y_i(1)}$ such that $Y_i = Y_i(W_i)$.
Then, we can identify a weighted treatment effect $\tau_c = \mathbb E\{Y_i(1) - Y_i(0) \mid Z_i = c\}$ via
\begin{equation}
\label{eq:continuity}
\tau_{c} = \lim_{z \downarrow c} \mathbb E(Y \mid Z = z) - \lim_{z \uparrow c}\mathbb E(Y \mid Z = z),
\end{equation}
provided that the conditional response functions \smash{$\mu_{(w)}(z) = \mathbb E\{Y(w) \mid Z = z\}$} are continuous.
As we further explain in Section~\ref{subsec:related_work}, if we are willing to posit quantitative smoothness bounds on
\smash{$\mu_{(w)}(\cdot)$}, then we
can use this continuity-based argument to derive confidence intervals for $\tau_c$ with well understood
asymptotics.

Despite its appeal and simple formulation,
the continuity-based approach to regression discontinuity inference
does not satisfy the criteria for
credible causal inference as outlined by \citet{rubin2008objective}. These criteria include  outcome-free design and modeling randomness in the assignment mechanism rather than modeling the outcomes. \citet{rubin2008objective} advocates for these principles to approximate the ideal of a design-based analysis of randomized controlled trials following \citet{neyman1923applications} and \citet{rubin1974estimating}. 
In contrast, the formal guarantees provided by  continuity-based regression discontinuity analyses often take smoothness
of \smash{$\mu_{(w)}(\cdot)$} as a primitive without explicitly  modeling effective randomization in the assignment mechanism. While continuous measurement error in (or imprecise control of) the running variable implies continuity of \smash{$\mu_{(w)}(\cdot)$} \citep{lee2008randomized}, this result is not used in estimation and inference.

Here we propose a new approach to regression discontinuity inference that goes back to the above qualitative argument used to justify regression discontinuity designs: Our approach
directly exploits effectively random treatment assignment induced 
by noise in the running variable $Z_i$.
Formally, we assume the existence of a latent variable $U_i$, and that the variation
in the running variable $Z_i$ around $U_i$ is known and exogenous. For example, revisiting our educational setting, we can take $U_i$ to be a measure of the student's true ability; then the test score $Z_i$ is
a noisy measurement of $U_i$ with well-documented psychometric properties. Likewise, in a medical setting, the running
variable $Z_i$ may be a measurement of an underlying condition $U_i$ (e.g., CD4 counts); such diagnostic measurements
often have well-studied test--retest reliability. In both cases, it is plausible that the measurements $Z_i$ are independent of
relevant potential outcomes conditional on the underlying quantity $U_i$.

Our main result is that, under our assumptions, we can estimate weighted treatment effects that correspond to the effects of realistic changes to the existing treatment assignment rule.
We then propose a practical approach to estimation and inference in regression discontinuity
designs that builds on this result. 
Our approach is conceptually appealing, offers transparency on the  key assumptions driving inference (noise-induced randomization and knowledge of noise mechanism), and allows for inference
of policy-relevant estimands beyond~\eqref{eq:continuity}.

Our approach is motivated by settings where the researcher has limited understanding of the response variable $Y_i$ and the causal mechanism connecting $Y_i$ and the treatment $W_i$, but has substantive knowledge about the running variable $Z_i$. In this sense, our framework is akin to the model-X knockoff framework for controlled variable selection \citep{candes2018panning}, which posits knowledge of the entire covariate distribution to facilitate inference of a poorly understood response variable conditionally on well-understood covariates.

We emphasize that, while this noise-induced randomization approach applies to many settings of interest,  it
does not apply to all regression discontinuity designs. Some running variables are not readily interpretable as
having measurement error or other exogenous noise; or we may not have a-priori information on the distribution
of this noise. For example, numerous studies have used geographic boundaries as discontinuities
\citep{keele2014geographic, rischard2018bayesian}, but it would be questionable to model the location of a
household in space as having meaningful measurement error. 
Likewise, analyses of close elections, which are a central example of regression discontinuity designs in political science and
economics \citep{caughey2011elections,lee2008randomized}, may not allow for a natural noise model for $Z_i$ that
would arise from, e.g., noisy counting of  ballots, though perhaps there are other sources of exogenous noise \citep[e.g., weather, ][]{gomez2007republicans,cooperman2017randomization}.
These considerations call attention
to the limits of the proposed approach, but also highlight a difference in the foundational assumptions required for identification, estimation, and inference in regression discontinuity designs with a noisy running variable versus the assumptions required when the running variable is noiseless.

\section{A latent variable model for regression discontinuity designs}
\label{subsec:latent}

We consider the classical sharp regression
discontinuity design with potential outcomes:
\begin{assumption}[Sharp regression discontinuity design]
\label{assu:rdd}
There are $i = 1, \, ..., \, n$ independent, identically distributed samples
$\cb{Y_i(0), \, Y_i(1), \, Z_i} \in \RR^3$ and a cutoff $c \in \RR$ such that units 
are assigned treatment via $W_i = 1(Z_i \geq c)$. For each sample,
we observe pairs $\cb{Y_i, \, Z_i}$ with $Y_i = Y_i(W_i)$.
\end{assumption}
Our approach requires domain-specific knowledge about the distribution of the running variable.
\begin{assumption}[Noisy running variable]
\label{assu:noise}
There is a latent variable $U_i$ with (unknown) distribution $G$ such that $Z_i \mid U_i \sim p(\cdot \mid U_i)$ for a known conditional density $p(\cdot \mid \cdot)$ with respect to a measure $\lambda$.
\end{assumption}
\noindent
Qualitatively, we interpret the latent variable $U_i$ in Assumption \ref{assu:noise} as a true measure of
the property we want to use for treatment assignment, e.g., $U_i$ could capture ability in an educational setting
or health in a medical one. 
The observed running variable $Z_i$ is then a noisy realization of $U_i$. The more noise there is in the running variable, the greater the effective randomization becomes, making our task easier. The assumption is flexible and can accommodate a wide range of noise models, including heteroskedastic noise and discrete running variables.

\begin{example}[Gaussian measurement error]
  \label{example:gaussian}
One common example of measurement error we consider is Gaussian, i.e., $Z_i \mid U_i \sim \nn\p{U_i, \, \nu^2}$ for $\nu > 0$. In the limit $\nu \to \infty$, treatment assignment is purely random and we get a  randomized controlled trial. Conversely, as $\nu \to 0$, there is no random noise and no randomness in the treatment assignment, so it is not possible to have inference justified by effective noise-induced randomization.

If in the true noise process \smash{$Z_i \mid U_i' \sim \nn\p{U_i', \, \nu_i^2}$}, where $U_i'$ is a latent variable that may be correlated with $\nu_i$, then Assumption~\ref{assu:noise}
holds if we posit a homoskedastic noise model \smash{$Z_i \mid U_i \sim \nn\p{U_i, \, \hnu^2}$} so long as $\nu_i \geq \hnu$ almost surely. The reason is that we may decompose $Z_i = U_i' + \eta_i = U_i' + \delta_i + \varepsilon_i = U_i + \varepsilon_i$, where $\eta_i \sim \mathcal{N}(0, \nu_i^2)$, $\delta_i  \sim \mathcal{N}(0, \nu_i^2 - \hnu^2)$ is independent of $\varepsilon_i \sim \mathcal{N}(0, \hnu^2)$, and $U_i = U_i' + \delta_i$.

Thus, the Gaussian measurement error assumption also accommodates heteroskedasticity and remains valid if we use a conservative lower bound on measurement error variance. (However, underestimating the measurement error
will result in a loss of power, since underestimating the measurement error reduces the number of units in the effective overlap region where treatment and control assignments are both a-priori plausible.)

\end{example}

\begin{example}[Binomial measurement error]
Assumption~\ref{assu:noise} also accommodates discrete running variables, such as $Z_i \mid U_i \sim \text{Binomial}(K, U_i)$ for some $K \in \mathbb N$.
\end{example}

Assumption~\ref{assu:noise} does not impose any restriction on $G$, which is a property of the studied population.  While precise knowledge about the noise distribution $p(z \mid u)$ is a strong requirement, it enables us to obtain strong results (valid causal estimates justified via a form of effective randomization).
In some applications, knowledge of the noise distribution may be available from test--retest data, prior modeling of item-level responses to tests, a physical model for the measurement device, biomedical knowledge, or direct control by the experimenter, e.g., in applications involving differential privacy~\citep{dwork2013algorithmica}. Any posited noise model should be carefully scrutinized since the credibility of the noise-induced randomization approach depends on the credibility of the noise model.

We also require for the additional noise to be exogenous.
We formalize this requirement in terms of an unconfoundedness condition following
\citet{rosenbaum1983central}.
\begin{assumption}[Exogeneity]
\label{assu:exogenous}  
The noise in $Z_i$ is exogenous, i.e., $\sqb{\cb{Y_i(0), \, Y_i(1)} \indep Z_i} \mid U_i$.
\end{assumption}

\begin{figure}
  \centering
  \includegraphics{Tikz/dag.tikz}
  \caption{Graphical illustration of the sharp regression discontinuity design with a noisy running variable. $U$ is an unobserved latent variable with unknown distribution $G$ and $Z$ is the running variable with known density $p(\cdot \mid U)$ conditionally on $U$. Treatment is assigned as $W = 1(Z \geq c)$ for a known cutoff $c$ and $Y=Y(W)$ is the observed response.}
  \label{fig:dag}
\end{figure}

An implication of Assumption~\ref{assu:exogenous} is that
\begin{equation}
\label{eq:cond_response}
\mathbb E(Y_i \mid U_i, \, Z_i) = \alpha_{(W_i)}(U_i), \ \ \ \ \alpha_{(w)}(u) = \mathbb E\{Y_i(w) \mid U_i = u\},
\end{equation}
where the \smash{$\alpha_{(w)}(u)$} are the response functions for the potential outcomes
conditionally on the latent variable $u$. 
Following \citet{frangakis2002principal} we can think of $u$ as indexing over unobserved principal strata; see also \citet{heckman2005structural}.

A graphical illustration of our assumptions is presented in Fig.~\ref{fig:dag}. In view of Assumptions~\ref{assu:noise} and \ref{assu:exogenous}, the key argument for our strategy is captured by the following proposition.
\begin{proposition}
\label{prop:balance_response}
Let $\gamma_+(\cdot), \gamma_-(\cdot)$ be functions of $z$ with $\gamma_+(z) = 0 \text{ for } z < c$, $\gamma_-(z) = 0 \text{ for } z\geq c$.
Suppose that Assumptions~\ref{assu:rdd}--\ref{assu:exogenous} hold and that $\mathbb E(Y^2_i), \mathbb E\{\gamma_-(Z_i)^2\}, \mathbb E\{\gamma_+(Z_i)^2\}<\infty$. Then:

\begin{equation}
\label{eq:hkernel_def}
\begin{aligned}
&\mathbb E\{\gamma_+(Z_i) Y_i\} = \mathbb E\{\hkernel{U_i}{\gamma_+}\alpha_{(1)}(U_i)\},\;\;\mathbb E\{\gamma_-(Z_i) Y_i\} = \mathbb E\{\hkernel{U_i}{\gamma_-}\alpha_{(0)}(U_i)\}, \\ 
&\text{ where } \,\,\, \textstyle \hkernel{u}{\gamma} = \int \gamma(z) p(z \mid u)d\lambda(z).
\end{aligned}
\end{equation} 
\end{proposition}
We will apply this result by choosing functions $\gamma_+, \gamma_-$ and then averaging the response $Y_i(1)$ of treated units with weights $\gamma_+(Z_i)$ and the response $Y_i(0)$ of control units with weights $\gamma_-(Z_i)$. While there is no overlap between treated and control units in a sharp regression discontinuity design in terms of the running variable $Z_i$, Proposition~\ref{prop:balance_response} establishes that by weighting treated units by $\gamma_+$ and control units by $\gamma_-$ we may achieve balance in the latent variable, if $\hkernel{\cdot}{\gamma_+} \approx \hkernel{\cdot}{\gamma_-}$.

\section{Related work}
\label{subsec:related_work}

As discussed above, the dominant approach to inference in regression discontinuity designs is via continuity-based
arguments that build on \eqref{eq:continuity}. Perhaps the most popular continuity-based approach is to use local
linear regression to estimate the treatment effect~\eqref{eq:continuity} at $Z_i = c$.
This approach can be used for valid estimation and inference of $\tau_c$ provided the functions $\mu_{(w)}(z) = \mathbb E\{Y_i(w) \mid Z_i=z\}$ are smooth
and the local linear regression bandwidth
decays at an appropriate rate; the rate of convergence of $\htau_c$ and appropriate choice of bandwidth
depend on the degree of smoothness assumed. Notable results in this line of work, covering topics such as robust confidence intervals
and data-adaptive bandwidth choices, include \citet{armstrong2016simple},
\citet{calonico2014robust} 
and \citet{imbens2011optimal}, 
as well as 
Bayesian approaches~\citep{branson2019nonparametric,geneletti2015bayesian}.
More recently, extensions have been considered to the continuity-based approaches that improve over local linear regression by directly exploiting the assumed smoothness
properties of $\mu_{(w)}(\cdot)$. Under the assumption that $\mu_{(w)}(\cdot)$ belongs to a convex class, e.g., \smash{$|\mu''_{(w)}(z)| \leq B$} for
all $z \in \RR$, \citet{armstrong2018optimal} and \citet{imbens2019optimized} use numerical convex optimization to
derive minimax linear estimators of \smash{$\tau_c$}.

One alternative approach to inference in regression discontinuity designs, which
\citet{cattaneo2015randomization}, \citet{li2015evaluating} and \citet{mattei2016regression} refer to as
local randomization inference, starts by positing a non-trivial interval $\mathcal{I}$ with $c \in \mathcal{I}$,
such that
\begin{equation}
\label{eq:pure_randomization}
\sqb{Z_i \indep \cb{Y_i(0), \, Y_i(1)}} \; \mid \; \cb{Z_i \in \mathcal{I}}.
\end{equation}
They then focus on the subset of units with $Z_i \in \mathcal{I}$, and perform classical randomized study inference
on this subset. Unlike continuity-based analysis, this approach is design-based in the sense of
\citet{rubin2008objective}.
In practice, however, the assumption \eqref{eq:pure_randomization} is often unrealistic and limits
the applicability of methods relying on it \citep{sekhon2017interpreting}. A testable implication of \eqref{eq:pure_randomization} is
that $\mu_{(w)}(z)$ should be constant over $\mathcal{I}$ for both $w = 0$ and $1$, but this structure rarely
plays out in the data. One may try to fix this issue by de-trending outcomes and assuming \eqref{eq:pure_randomization} on the residuals~\citep{sales2020limitless}; however, such an approach relies on correct specification of the trend removal, and is thus no longer justified by randomization. Furthermore, it is not clear how to choose the interval $\mathcal{I}$ used in \eqref{eq:pure_randomization} via the types of methods
typically used for regression discontinuity inference. There's no data-driven way of discovering an interval $\mathcal{I}$ over which
\eqref{eq:pure_randomization} holds that is itself justified by randomization; conversely, if the interval $\mathcal{I}$ is known
a-priori, then the problem collapses to a basic randomized controlled trial where the regression discontinuity structure is not used for inference. 

The idea that explicit structural modeling is valuable for causal inference has a long tradition in economics,
going back to \citet{roy1951some} and \citet{heckman1979sample}, with recent developments by e.g.,
\citet{heckman2005structural}, \citet{brinch2017beyond} and \citet{mogstad2018using}.
At a high level, our work can be seen as connecting this tradition to the regression discontinuity design, and demonstrating how
structural assumptions enable inference of policy-relevant causal estimands.

Knowledge of the presence of measurement error (or other noise) in running variables is often mentioned
\citep{bor2014regression,bor2017treatment,harlow2020impact,lee2008randomized}, yet this side-information is typically not directly used for inference. In a rare quantitative use of information about measurement error, \citet{fraga2016examining} use margin of error statistics provided by the Census Bureau for the fraction or size of a voting-aged population that has limited English proficiency; they report some analyses using only units that are within a 90\% margin of error of the cutoff.

Closer to our approach, \citet{rokkanen2015exam} considers the regression discontinuity design 
under Assumptions~\ref{assu:noise} and~\ref{assu:exogenous}. 
Instead of assuming prior knowledge of the noise distribution $p(\cdot \mid u$),
\citet{rokkanen2015exam} assumes that for each unit we observe at least two noisy measurements
$Z_i', Z_i''$ of the underlying latent variable $U_i$ in addition to the running variable $Z_i$.
While \citet{rokkanen2015exam} provides conditions for the nonparametric identification of $\alpha_{(w)}(\cdot)$ in~\eqref{eq:cond_response} and consequently of treatment effects, the estimation and inference strategy posits strong parametric assumptions,
namely joint normality of $(U_i, Z_i, Z_i', Z_i'')$ and linearity of $\alpha_{(w)}(u)$ as a function of $u$.
In contrast, we assume knowledge of the noise distribution through, e.g., biomedical knowledge or test--retest data, however we impose no parametric restrictions on $G$ and $\alpha_{(w)}(u)$. 
Furthermore, we develop a practical and intuitive method for estimation and inference, that provides valid coverage
even when treatment effects are only partially identified (e.g., when $p(\cdot \mid u)$ is finitely supported).

Our results are also connected to research on treatment effect estimation under
biased or risk-based allocation
\citep{robbins1989estimating, robbins1991estimating, finkelstein1996partB}
motivated from an empirical Bayes 
interpretation of the noise model in Assumption~\ref{assu:noise}.
For example, \citet{robbins1989estimating} study treatment
effect estimation under what effectively amounts to our Assumptions \ref{assu:noise} and \ref{assu:exogenous} with Gaussian noise, \smash{$Z_i \mid U_i \sim \nn(U_i, \, \nu^2)$}, and control potential outcomes linked to $U_i$ via an additive shift, \smash{$\alpha_{(0)}(u) = \mathbb E\{Y_i(0) \mid U_i = u\} = u + s$} for \smash{$s \in \RR$}.
These assumptions on $\alpha_{(0)}(u)$ are motivated by settings where measurements of the same quantity function as both the running variable and the outcome, such as the application in \citet{finkelstein1996partB} where patients with high cholesterol are given a drug to lower cholesterol, and we are interested in the drug's effectiveness. However, such assumptions on control outcomes
are not
appropriate in the examples considered in this paper. Thus, while this line of work presents a notable yet largely overlooked chapter in the history of regression discontinuity designs~\citep{cook2008waiting}, it does not provide a methodological baseline for our approach.

Finally, we contrast our setup with a line of work that studies the regression discontinuity design
when the running variable is unobserved, and instead a noisy measurement thereof is observed
\citep[e.g.,][]{davezies2017regression, dong2021can};
see the causal diagram in Supplementary Fig.~\ref{fig:dag_deconv} for an illustration.
Identification becomes subtle and estimation can be difficult because of the difficulties of nonparametrics with measurement error~\citep{meister2009deconvolution}. 
Instead, we use measurement error as our identifying assumption; 
the noise in our setup is beneficial for our estimation strategy rather than a barrier (and we observe the running variable).

\section{Ratio-form estimators and weighted treatment effects}
\label{sec:identification}
\subsection{Preliminaries}
Motivated by Proposition~\ref{prop:balance_response}, we consider ratio-form estimators,
\begin{equation}
\label{eq:weighted}
\htau_\gamma = \hmu_{\gamma,+}-\hmu_{\gamma,-},\;\;  \hmu_{\gamma,+}= \frac{\sum_i \gamma_+(Z_i) Y_i}{\sum_{i} \gamma_+(Z_i)},\;\; \hmu_{\gamma,-} = \frac{\sum_{i} \gamma_-(Z_i) Y_i}{\sum_{i} \gamma_-(Z_i)},
\end{equation}
where $\gamma_+, \gamma_-$ are pre-specified weighting functions such that $\gamma_+(z) = 0 \text{ for } z < c$, $\gamma_-(z) = 0 \text{ for } z\geq c$. The class \eqref{eq:weighted} is a broad and intuitive class of estimators that includes, for example, the difference-in-means of units that are close to the cutoff (with the choice $\gamma_+(z) = \ind\{z \in [c, c+h]\}$ and $\gamma_-(z) = \ind\{z \in [c-h, c)\}$ for $h>0$). 

Our goal is to conduct inference for weighted treatment effects,
\begin{equation}
\label{eq:specific_weighted_target}
\tau_w = \textstyle \int [w(u)\,/\,\mathbb E_G\{w(U_i)\}] \tau(u) \, dG(u),\;\;\; w(\cdot) \geq 0,
\end{equation}
where $\tau(u)$ is the conditional average treatment effect (CATE) of the stratum with $U_i = u$,
\begin{equation*}
\label{eq:strat}
\tau(u) = \mathbb E\{Y_i(1) - Y_i(0) \mid U_i = u\} = \alpha_{(1)}(u) - \alpha_{(0)}(u),
\end{equation*}
and $w(\cdot)$ is a latent weighting (i.e., $w(\cdot)$ assigns weight to the latent $U$) chosen by the analyst.

In the following, we take $\gamma_+, \gamma_-$ as pre-specified by the researcher and seek to understand how to use the point estimate $\htau_{\gamma}$ in~\eqref{eq:weighted} to form valid confidence intervals for $\tau_w$ in~\eqref{eq:specific_weighted_target} by accounting for potential bias. In Section~\ref{sec:weighted_opt}, we make a concrete recommendation for choosing $\gamma_+, \gamma_-$.

\subsection{An asymptotic bias decomposition}
\label{subsec:weighted_effects}

We first derive the asymptotic limit of $\htau_{\gamma}$ with fixed $\gamma_+(\cdot), \gamma_-(\cdot)$ given $n$ i.i.d. copies of $(U_i, Z_i, Y_i(0), Y_i(1))$ satisfying Assumptions~\ref{assu:rdd}-\ref{assu:exogenous}. 

\begin{theorem}
\label{theo:consistency}
Suppose that Assumptions~\ref{assu:rdd}-\ref{assu:exogenous} hold, that $\mathbb E\{\gamma_+(Z_i)^2\}, \, \mathbb E\{\gamma_-(Z_i)^2\},\, \mathbb E(Y_i^2) < \infty$, and that $\mathbb E\{\gamma_+(Z_i)\},\, \mathbb E\{\gamma_-(Z_i)\}>0$. Then as $n\to \infty$, $\htau_\gamma - \taugamma\, \to \, 0$ in probability, where:
\begin{equation}
\label{eq:tau_gamma}
\begin{aligned}
\taugamma = \mu_{\gamma,+}-\mu_{\gamma,-},\;\;\mu_{\gamma,+}=\frac{\mathbb E\{\hkernel{U_i}{\gamma_+}\alpha_{(1)}(U_i)\}}{\mathbb E\{\hkernel{U_i}{\gamma_+}\}},\;\; \mu_{\gamma,-} = \frac{\mathbb E\{\hkernel{U_i}{\gamma_-}\alpha_{(0)}(U_i)\}}{\mathbb E\{\hkernel{U_i}{\gamma_-}\}}.
\end{aligned}
\end{equation}
\end{theorem}
In view of Theorem~\ref{theo:consistency} and the definition of $\tau_w$ in~\eqref{eq:specific_weighted_target}, we derive an asymptotic decomposition of the bias in estimating $\tau_w$ through $\htau_{\gamma}$:
\begin{corollary}
\label{coro:bias_decomposition}
Under the conditions of Theorem~\ref{theo:consistency}, the asymptotic bias $\taugamma - \tau_w$ decomposes as follows (making the dependence on $\gamma_{\pm},\, \tau_w$ and $\alpha_{(0)}(\cdot),\, \tau(\cdot),\,G$ explicit):
\begin{equation*}
\begin{aligned}
\operatorname{Bias}\{\gamma_{\pm},\, \tau_w; \, \alpha_{(0)}(\cdot),\, \tau(\cdot),\,G\}    &= \;  \underbrace{\int\sqb{\frac{\hkernel{u}{\gamma_+}}{\mathbb E_G\{\hkernel{U_i}{\gamma_+}\}} - \frac{\hkernel{u}{\gamma_-}}{\mathbb E_G\{\hkernel{U_i}{\gamma_-}\}}} \alpha_{(0)}(u)\, dG(u)}_{{\color{gray} \text{Confounding bias}} } \\
 &\;\;\;\;\;\;\; + \underbrace{\int \sqb{\frac{\hkernel{u}{\gamma_+}}{\mathbb E_G\{\hkernel{U_i}{\gamma_+}\}} - \frac{w(u)}{\mathbb E_G\{w(U_i)\}}} \tau(u)dG(u)}_{\color{gray} \text{Heterogeneity bias}}.
\end{aligned}
\end{equation*}
\end{corollary}
The bias decomposes into two terms. The first term (confounding bias) describes how well we are balancing units through their latent variable $u$ and will be small if $\hkernel{\cdot}{\gamma_+} \approx \hkernel{\cdot}{\gamma_-}$. The second term, which we call heterogeneity bias, is equal to zero when the conditional average treatment effect $\tau(u)$ is constant as a function of $u$, or when $\hkernel{u}{\gamma_+}=w(u)$ for all $u$.

\subsection{Examples of weighted treatment effects}

We now provide examples of estimands that may be expressed as weighted treatment effects~\eqref{eq:specific_weighted_target}.

\begin{example}[Regression discontinuity parameter]
The standard regression discontinuity parameter
$\tau_c$, defined in \eqref{eq:continuity}, may be written as in~\eqref{eq:specific_weighted_target}. By Bayes' rule,
\begin{equation}
\label{eq:RDparam}
\tau_c = \mathbb E\{Y_i(1) - Y_i(0) \mid Z_i=c\} = \mathbb E\{\tau(U_i) \mid Z_i=c\} = \textstyle\int \tau(u)p(c \mid u)dG(u) \big/  f(c),
\end{equation}
where \smash{$f(c)=f_G(c) = \smallint p(c \mid u) dG(u)$} is the density of the running variable $Z_i$ at the cutoff $c$. Thus, the representation from~\eqref{eq:specific_weighted_target} holds with $w(u) = p(c \mid u)$ and $\mathbb E\{w(U_i)\}=f(c)$. 

Another closely related target is $\tau_{c'}$ as defined in \eqref{eq:RDparam}, but for some other value $c' \neq c$ of the
running variable. This estimand also fits in our setting, with $w(u) = p(c' \mid u)$ and $\mathbb E\{w(U_i)\}=f(c')$.
Conceptually, estimating $\tau_{c'}$ away from $c$ involves extrapolating treatment effects away from the
cutoff \citep{angrist2015wanna,dong2015identifying,rokkanen2015exam}.
\end{example}

\begin{example}[Changing the cutoff]
As argued in \citet{heckman2005structural}, in many settings we may be most interested in evaluating
the effect of a policy intervention. One  case of a policy intervention involves changing the eligibility
threshold, i.e., standard practice involves prescribing treatment
to subjects whose running variable crosses $c$, but we are now considering changing this cutoff to a new
value $c' < c$.
For example, in a medical setting, we may consider lowering the severity threshold at which
we intervene on a patient. In this case, we need to estimate the average treatment effect $\tau_{\pi}$ of patients affected by the treatment which, in this case, amounts to:
\begin{equation}
\label{eq:change_cutoff}
\tau_\pi = \mathbb E\{\tau(U_i)\mid Z_i \in [c', c)\} = \textstyle \int_{[c',c)} \int \tau(u)p(z \mid u) dG(u) d\lambda(z)\big/ \int_{[c',c)} dF(z),
\end{equation}
where $F = F_G$ is the marginal $Z$-distribution (i.e., the distribution with $d\lambda$-density $f=f_G$). By Fubini's theorem, $\tau_\pi$ can be written as in~\eqref{eq:specific_weighted_target} with 
$w(u)= \smallint_{[c',c)} p(z \mid u) d\lambda(z)$.
\end{example}
In Supplement~\ref{sec:measurement_error_estimand}, we also consider an estimand motivated by a policy intervention that involves reducing measurement error in the running variable.

\section{Bias-aware confidence intervals}
\label{sec:inference}
\subsection{General inferential strategy}

In the previous section, we discussed the asymptotic limit of the ratio-form estimator in~\eqref{eq:weighted} and the bias in estimating weighted treatment effects in regression discontinuity designs. To make use of such an estimator in practice,
however, we also need to understand its sampling distribution and to control the bias. In this section, we describe our approach to inference.

We make the following additional assumption:
\begin{assumption}[Bounded response]
\label{assu:bounded}  
The response $Y_i$ is bounded, $Y_i \in [0,1]$.
\end{assumption}
We start by studying the asymptotic distribution of the ratio-form estimator in~\eqref{eq:weighted}.
We treat $\gamma_+, \gamma_-$ as deterministic but (in contrast to Theorem~\ref{theo:consistency}) allow them to vary with $n$, i.e.,
\smash{$\gamma_+ = \gamma_+^{(n)}$} and \smash{$\gamma_- = \gamma_-^{(n)}$}. Our first formal result is the following central limit theorem.

\begin{theorem}[Asymptotic normality of ratio-form estimators]
\label{theo:main_clt}
Suppose that Assumptions~\ref{assu:rdd}-\ref{assu:bounded} hold and that $\inf_z \operatorname{Var}(Y_i \mid Z_i=z) >0$. Further suppose that \smash{$\gamma_+^{(n)}$} and \smash{$\gamma_-^{(n)}$} are deterministic,
and that there exist $\beta \in (0,1/2)$, $C,C'>0$ such that for all $n$ large enough:
\begin{equation}
\label{eq:regular_weighting_kernel}
 \sup_{z} |\gamma_{\diamond}^{(n)}(z)| < C n^{\beta} \mathbb E\{\gamma_{\diamond}^{(n)}(Z_i)\},\;\; \sup_u |\hkernel{u}{\gamma_{\diamond}^{(n)}}| < C'\mathbb E\{\gamma_{\diamond}^{(n)}(Z_i)\},\;\;\diamond \in \cb{+,-}.
\end{equation}
Then, $\hat{\tau}_{\gamma} = \hat{\tau}_{\gamma^{(n)}}$ is asymptotically normal, 
$\sqrt{n}\p{\hat{\tau}_{\gamma}-\taugamma} \,\big/\, {\sqrt{V_{\gamma}}} \Rightarrow \nn\p{0, \, 1},$
where $\taugamma$ is defined in~\eqref{eq:tau_gamma} and
$V_\gamma = \mathbb E\{\gamma^2_+(Z_i) (Y_i - \mu_{\gamma,+})^2\}/ \mathbb E\{\gamma_+(Z_i)\}^2  + \mathbb E\{\gamma^2_-(Z_i) (Y_i - \mu_{\gamma,-})^2\} /\mathbb E\{\gamma_-(Z_i)\}^2$. 
\end{theorem}
The assumption on $\gamma_+,\gamma_-$ is satisfied by the weighting functions proposed in Section~\ref{sec:weighted_opt} and other choices. For example, the local difference-in-means estimator with $\gamma_+(z) = 1\{z \in [c, c+h_n]\}$, $\gamma_-(z) = \ind\{z \in [c-h_n, c)\}$ meets the assumption when $h_n^{-1} = O(n^{\beta})$ for $\beta \in (0,1/2)$, $\lambda$ is the Lebesgue measure and $\cb{p(\cdot \mid u)}_u$ are uniformly bounded and equicontinuous at $c$.

To construct confidence intervals for  $\tau_w$
in~\eqref{eq:specific_weighted_target}, we first estimate the asymptotic variance \smash{$V_\gamma$}.
\begin{proposition}
\label{prop:vgamma_consistent}
Let $\hmu_{\gamma,+}, \hmu_{\gamma,-}$ be defined as in~\eqref{eq:weighted}.
Under the assumptions of Theorem~\ref{theo:main_clt}, $V_{\gamma}$ can be consistently estimated with
the following plug-in estimator: \smash{${\hV_{\gamma}} \,\big/\,{V_{\gamma}} = 1 + o_{\mathbb P}(1)$} for
\begin{equation}
\label{eq:plugin_variance}
\hV_{\gamma} = \frac{\sum_{i} \gamma_+(Z_i)^2(Y_i - \hmu_{\gamma,+})^2}{n\big\{\frac{1}{n}\sum_{i} \gamma_+(Z_i)\big\}^2} + \frac{\sum_i \gamma_-(Z_i)^2(Y_i - \hmu_{\gamma,-})^2}{n\big\{\frac{1}{n}\sum_{i} \gamma_-(Z_i)\big\}^2}.
\end{equation}
\end{proposition}
Second, we account for the potential bias \smash{$|b_{\gamma}| = |\taugamma - \tau_w|$}. We do not assume negligible bias
(i.e., undersmoothing) and accommodate settings wherein treatment effects are only partially identified,
and bias does not decay to zero even asymptotically~\citep[Section II.A]{imbens2019optimized} (e.g., when \smash{$Z_i \mid U_i$} has a binomial distribution). To do so, we derive an upper bound \smash{$\hB_{\gamma}$} for the
bias $|b_{\gamma}|$.
A challenge is that the expectations in Corollary~\ref{coro:bias_decomposition}
involve integrals over the
latent variable $U_i$ and the unknown functions $G$, $\tau(\cdot)$ and $\alpha_{(0)}(\cdot)$. Taking a clue from~\cite{ignatiadis2019bias}, we bound the worst-case bias over any data-generating distribution consistent with the observed data for the running variable $Z_i$. Define the marginal distribution function of $Z_i$ (marginalizing over $U_i \sim G$), $F_G(t)=\smallint 1(z \leq t) \smallint p(z\mid u)dG(u)d\lambda(z)$, and let $\mathcal{G}_n$ be the class of latent variable distributions with marginal distribution $F_G$ inside the Kolmogorov-Smirnov band~\citep{massart1990tight} centered at the empirical distribution $\widehat{F}_n(t) = \sum_{i=1}^n 1(Z_i \leq t)/n$,
\begin{equation*}
\label{eq:calG}
\mathcal{G}_n = \cb{G \text{ distrib.}: \; \sup_{t \in \mathbb R}| F_G(t) - \widehat{F}_n(t)| \leq  \sqrt{\log\p{2/\alpha_n}/{(2n)}}},\; \alpha_n = \min\big\{0.05, n^{-\frac{1}{4}}\big\}.
\end{equation*}
We also consider a sensitivity model for treatment effect heterogeneity. 
For $M \in [0,\,1]$, we let
\begin{equation}
\label{eq:sensi}
\mathcal{T}_M = \cb{\tau(\cdot) \, : \, \tau(u) = \bar{\tau} + \Delta(u) \;\text{ for } \bar{\tau} \in \RR \text{ and } \Delta(\cdot) \text{ s.t. }\abs{\Delta(u)} \leq M}.
\end{equation}

\noindent Above, $\mathcal{T}_0$ consists of all conditional average treatment effect (CATE) functions $\tau(\cdot)$ that are constant as a function of $u$.
Under Assumption~\ref{assu:bounded},
$\mathcal{T}_1 = \cb{\text{all CATE functions } \tau(\cdot)},\; \mathcal{T}_{1/2} \supset \cb{\text{all CATE functions }  \tau(\cdot) \geq 0 },$
and so the choice $M=1$ avoids imposing any additional assumptions on heterogeneity, while $M=1/2$ is a conservative choice under the monotonicity restriction $\tau(\cdot) \geq 0$.
\begin{proposition}
\label{prop:validbiasbound}
Suppose that $\tau(\cdot) \in \mathcal{T}_M$ and that we upper bound the bias \smash{$|b_{\gamma}| = |\taugamma - \tau_w|$} as,
\begin{equation}
\label{eq:biasfractional}
\hB_{\gamma,M} = \sup\cb{\abs{\operatorname{Bias}\{\gamma_{\pm},\, \tau_w; \, \alpha_{(0)}(\cdot),\, \tau(\cdot),\,G\}} \,:\, G \in \mathcal{G}_n,\,\alpha_{(0)}(\cdot) \in [0,1],\,\tau(\cdot) \in \mathcal{T}_M}.
\end{equation}
Then under the assumptions of Theorem~\ref{theo:main_clt}, it holds that \smash{$\mathbb{P}(|b_{\gamma}| \leq \hB_{\gamma,M}) \rightarrow 1$} as $n \rightarrow \infty$.
\end{proposition}
In Supplement~\ref{subsec:bias_computation} we explain how to compute this bound on the bias. Finally, we build confidence intervals for $\tau$ that are robust to estimation bias up to \smash{$\hB_{\gamma,M}$} following \citet{imbens2004confidence}, \citet{armstrong2018optimal}, and \citet{imbens2019optimized}.

\begin{corollary}[Valid confidence intervals]
\label{coro:valid_ci}
Suppose the assumptions of Theorem~\ref{theo:main_clt} hold, and that $\tau(\cdot) \in \mathcal{T}_M$. Then, $\liminf_{n \to \infty} \mathbb P( \tau_w \in \htau_{\gamma} \pm \ell_\alpha) \geq 1-\alpha$, where
\begin{equation}
\label{eq:CI}
\ell_\alpha = \min\cb{\ell \,:\, \mathbb P\big(\big|b + n^{-1/2}\hV_\gamma^{1/2} \tZ\big| \leq \ell\big) \geq 1-\alpha \text{ for all } \abs{b} \leq \hB_{\gamma,M}},
\end{equation}
with $\tZ$ a standard Gaussian random variable, and $\alpha \in (0,1)$ the significance level.
\end{corollary}

\subsection{Robustness to misspecification of treatment effect heterogeneity}
\label{sec:robustness}

Our method requires specifying the sensitivity model~\eqref{eq:sensi}. While one can adopt the unrestrictive model $\mathcal{T}_1$, we explore the robustness of our approach to misspecification of the sensitivity model: We suppose that $\tau(\cdot)$ is not constant as a function of $u$, yet we conduct inference using $\mathcal{T}_0$. In this case, our intervals attain the correct coverage for a convenience-weighted treatment effect.
\begin{corollary}[Valid  intervals for the convenience-weighted treatment effect]
\label{coro:valid_ci_het}
Assume the conditions from Theorem~\ref{theo:main_clt} are satisfied. Also suppose that $\tau(\cdot) \in \mathcal{T}_M$ but we construct confidence intervals as in~\eqref{eq:CI} using $M' \geq 0$ instead of $M$ (say, $M' <M$). Then these confidence intervals satisfy
$\liminf_{n \to \infty} \mathbb P( \tau_{h,+} \in \htau_{\gamma} \pm \ell_\alpha) \geq 1-\alpha$,
where 
$$\tau_{h,+} = \textstyle \int [\hkernel{u}{\gamma_+} \,/\,\mathbb E_G\{\hkernel{U_i}{\gamma_+} \}] \tau(u) \, dG(u).$$
\end{corollary}
The convenience-weighted treatment effect $\tau_{h,+}$ may be of interest if we are not directly interested in treatment heterogeneity
\citep{crump2009dealing,li2016balancing,imbens2019optimized,kallus2020generalized}. 
If we are interested in the null hypothesis of no treatment effects, $H_0: \tau(u) =0 \text{ for all }u,$ then we can form a valid test by forming confidence intervals for $\tau_w$ under the sensitivity model $\mathcal{T}_0$ and rejecting the null hypothesis when the resulting confidence interval does not include $0$. 

\subsection{Balancing the response functions in the absence of effective randomization}

In Supplement~\ref{sec:robustness_balancing}, we study the asymptotic bias for our method 
when there is no effective randomization at all, but we proceed pretending Assumptions~\ref{assu:noise} and~\ref{assu:exogenous} hold for a given $p(\cdot \mid \cdot)$. Our result requires a strong functional form assumption according to which $\mu_{(w)}(z) = \mathbb E\{Y_i(w) \mid Z_i=z\}$ is a linear combination of $p(z \mid u)/f(z)$, where $f$ is the $d\lambda$-density of $Z$. The result implies consistency and valid inference (no matter the noise model we posit)
when unbeknownst to us, $\mu_{(w)}(z)$ is constant as a function of $z$.

\section{Designing estimators via quadratic programming}
\label{sec:weighted_opt}

Given a choice of weighting functions $\gamma_{\pm}$ for \eqref{eq:weighted}, Section~\ref{sec:inference} provides a complete recipe for building valid confidence intervals justified
by the noise-induced randomization framework. For example, 
one could take weighting functions implied by various regression discontinuity estimators. Existing weighting functions $\gamma_{\pm}$, however, were not designed for our
framework, and so may not yield particularly short confidence intervals. Hence we turn to deriving weighting functions $\gamma_{\pm}$ with an eye towards making confidence
intervals obtained via Corollary \ref{coro:valid_ci} short.

Our (heuristic) strategy is to choose $\gamma_{\pm}$ by minimizing an approximate bound on the worst-case mean-squared error of the estimator in \eqref{eq:weighted}. Let $w(\cdot)$ be the latent weighting of the estimand~\eqref{eq:specific_weighted_target} and suppose we posit the sensitivity model $\mathcal{T}_M$. Furthermore, let $\bar{F}(\cdot)$ be a guess or estimate of the marginal distribution $F_G(\cdot)$ of $Z_i$ under Assumption~\ref{assu:noise} and let $\bar{w}(\cdot)$ be an estimate of the normalized latent weighting $w(\cdot)/\mathbb E_G\{w(U_i)\}$. We propose solving the following quadratic program (of which an appropriately discretized version can be solved using standard convex optimization software, e.g., MOSEK,~\citealp{mosek}):
\begin{subequations}
\label{eq:qp}
\begin{align}
\min_{\gamma_{\pm}(\cdot)} \quad  & \frac{1}{n} \cb{\textstyle  \int \gamma_-^2(z) \, d\bar{F}(z)  + \textstyle \int \gamma_+^2(z)  \,  d\bar{F}(z)}\;+ \; \p{t_1+ t_2}^2 \label{eq:simpleQP}
 \\
\textrm{s.t.} \quad &  \abs{\hkernel{u}{\gamma_+} - \hkernel{u}{\gamma_-}} \leq t_1,\;M \abs{\hkernel{u}{\gamma_{\diamond}}   - \bar{w}(u)}  \leq t_2 \text{ for } \diamond \in \cb{\pm} \text{ and all } u, \label{eq:simpleQP_bias}\\
&\textstyle \int \gamma_-(z)d\bar{F}(z) = 1, \;\;\;\; \textstyle  \int \gamma_+(z)d\bar{F}(z) = 1,  \label{eq:integrate_to_1}\\
&\gamma_-(z) = 0 \text{ for } z\geq c, \;\; \gamma_+(z) = 0 \text{ for } z < c, \label{eq:support} \\
& \abs{\gamma_{\diamond}(z)} \leq Cn^{\beta} \text{ for } \diamond \in \cb{\pm} \text{ and all } z. \label{eq:optimization_clt} 
\end{align}
\end{subequations}
In choosing $\bar{F}(\cdot)$ and $\bar{w}(\cdot)$, we make use of the structure
provided by Assumption \ref{assu:noise}, and estimate $G$ as $\bar{G}$ via nonparametric maximum likelihood~\citep{kiefer1956consistency} and then we let $\bar{F}(\cdot) = F_{\bar{G}}(\cdot)$ and $\bar{w}(\cdot) = w(\cdot)/\mathbb E_{\bar{G}}\{w(U_i)\}$.

The first term in~\eqref{eq:simpleQP} is a variance proxy for our estimator, motivated by the inequality $\operatorname{Var}(\gamma_{\diamond}(Z_i)Y_i) \leq \smallint \gamma_{\diamond}^2(z) \, dF(z)$ for $\diamond \in \cb{\pm}$. 
The second term, $(t_1 + t_2)^2$,  approximately bounds the worst-case bias. 
The bias is decomposed through the triangle inequality into  two terms resembling the bias decomposition of Corollary~\ref{coro:bias_decomposition};
$t_1$ in \eqref{eq:simpleQP_bias} seeks to bound the confounding bias and balances $\hkernel{\cdot}{\gamma_+}$ and $\hkernel{\cdot}{\gamma_-}$,
while $t_2$ seeks to bound the heterogeneity bias and balances $h$ with the normalized $w(\cdot)$.
Next, \eqref{eq:integrate_to_1} is a normalization constraint, and~\eqref{eq:support} enforces that $\gamma_+$  ($\gamma_-$) assigns weight only to treated (control) units. Constraint \eqref{eq:optimization_clt} 
ensures that no single observation has excessive influence (we omitted this constraint in our numerical implementation).

The following proposition shows that the weighting functions $\gamma_{\pm}$ derived from optimization problem~\eqref{eq:qp} satisfy the conditions of Theorem \ref{theo:main_clt} and thus enable valid inference.
\begin{proposition}
\label{prop:suff_weights}
Let $\gamma_{\pm}=\gamma_{\pm}^{(n)}$ be solutions to optimization problem~\eqref{eq:qp} for $M>0$, where $\bar{F}(\cdot), \bar{w}(\cdot)$ are guesses for $F_G(\cdot)$, $w(\cdot)/\mathbb E_G\{w(U_i)\}$ or estimates based on a held-out sample. Assume that $\bar{F}$ assigns non-trivial mass to $[c,\infty)$ and that $\bar{w}(\cdot)$ is bounded, i.e., there exists $k>1$  such that \smash{$\mathbb P\{ 1/k < \bar{F}([c,\infty)) < 1-1/k,\; \sup_u \abs{\bar{w}(u)} < k\} \to  1$} as $n\to \infty$ and that the expectation of $\gamma_{\pm}^{(n)}$ is asymptotically lower bounded by a strictly positive number, i.e.,
there exists $\delta>0$ such that \smash{$\mathbb P[ \smallint \gamma_{\diamond}^{(n)}(z) dF_G(z) > \delta, \, \diamond \in \cb{\pm}] \to  1$} as $n\to \infty$.
Then, the weighting functions $\gamma_{\pm}$ satisfy condition~\eqref{eq:regular_weighting_kernel} from Theorem~\ref{theo:main_clt} on an event $A_n$ with $\mathbb P(A_n) \to 1$ as $n\to \infty$.
\end{proposition}

In our implementation, we use all running variables $Z_i$ (but not the responses $Y_i$) to form estimates for $\bar{F}(\cdot)$ and $\bar{w}(\cdot)$; throughout our simulations we have not observed any undercoverage thereby. We summarize our approach in Algorithm \ref{algo:NIR}.

\begin{algorithm}
  \caption{Confidence intervals for treatment effects in regression discontinuity designs with
  noise-induced randomization (NIR).}
  \label{algo:NIR}
\KwIn{Samples $\{Z_i, Y_i,\, i = 1,\ldots,n\}$, cutoff $c$, sensitivity parameter $M \in [0,\,1]$ for the sensitivity model in~\eqref{eq:sensi}, estimand $\tau_w$ as in~\eqref{eq:specific_weighted_target}, significance level $\alpha \in (0,1)$.}
  1. Form a guess or estimate $\bar{F}$ of the marginal $Z$-distribution and $\bar{w}(\cdot)$ of the normalized latent weighting $w(\cdot)/\mathbb E_G\{w(U_i)\}$.\\
  2. Solve the minimax quadratic program in~\eqref{eq:qp} to get $\gamma_+, \gamma_-$.\\
  3. Form the point estimate $\htau_\gamma$ as in~\eqref{eq:weighted}.\\
  4. Estimate the asymptotic variance of $\htau_\gamma$ by \smash{$\hV_{\gamma}$ }as in~\eqref{eq:plugin_variance}.\\
  5. Estimate the worst-case bias \smash{$\hB_{\gamma}$} by~\eqref{eq:biasfractional}.\\
  6. Form bias-aware confidence intervals $\htau_{\gamma} \pm \ell_\alpha$ with $\ell_{\alpha}$ defined in~\eqref{eq:CI}.
\end{algorithm}

\section{Application: Antiretroviral Therapy Eligibility and Retention}
\label{sec:art}
\subsection{Background}
\label{subsec:art_background}

In this section, we apply our approach to a medical study. \citet{bor2017treatment}
study $11,306$ patients in South Africa (in 2011--2012) diagnosed with HIV aiming to understand whether
immediate initiation of antiretroviral therapy (ART) helps retain patients in the medical system. The response  $Y_i \in \cb{0,1}$ is an indicator of the $i$-th patient's retention, measured by the
presence of a clinic visit, lab test, or ART initiation 6 to 18 months after the initial HIV diagnosis. 

\begin{figure}
\centering
\includegraphics[width=\textwidth]{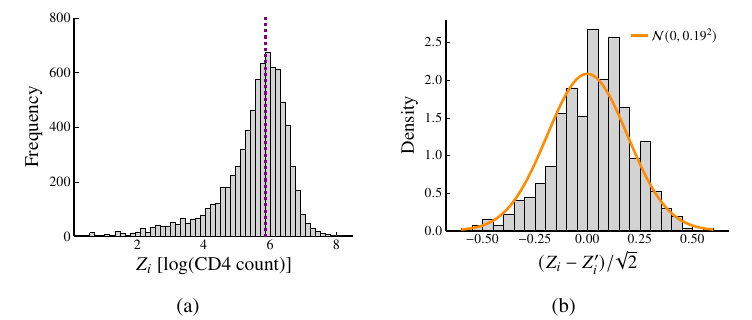}
\caption{CD4 counts ($\text{cells} /\mu L$) as a noisy running variable in a regression discontinuity analysis.
{(a)} Histogram of the running variable $Z_i$ in the dataset
of \citet{bor2017treatment}. {(b)} Differences $(Z_i-Z_i')/\sqrt{2}$ between repeated measurements
in the dataset of \citet{venter}, overlaid with a Gaussian probability density function.
}
\label{fig:bor_intro}
\end{figure}

According to health guidelines used in South Africa at the time, an HIV-positive patient should receive immediate ART if their measured CD4 count
was below $350\; \text{cells} /\mu L$ (a low CD4 count is indicative of poor immune function), lending itself to a natural regression discontinuity design for intention-to-treat effects. Figure~\ref{fig:bor_intro}(a) shows a histogram of the running variable $Z_i$, the log CD4 count (in cells/$\mu L$), with treatment cutoff $c = \log(350)$ denoted by a dashed line.

\citet{bor2017treatment} emphasize that CD4 count measurements are noisy; causes of this noise include instrument imprecision
and variability in the blood sample taken \citep[see, e.g.,][]{hughes1994within,wade2014multicenter}.
They then use the existence of such noise to qualitatively argue that treatment $W_i = \ind(Z_i < c)$ is effectively random close
to the cutoff $c$, thus strengthening the credibility of the regression discontinuity analysis. 

 Here, we seek an approach to estimating the effect of ART on retention that is driven
by the effective treatment randomization provided by the measurement error in $Z_i$ and (approximate) knowledge of the noise mechanism. To this end, we start by modeling this measurement error. \citet{venter} provide pairs of repeated
measurements \smash{$Z_i, \, Z_i'$} of the log CD4 count on 553 individuals (with measurements taken in the same laboratory).
Figure~\ref{fig:bor_intro}(b) compares a histogram of the normalized differences \smash{$(Z_i-Z_i')/\surd{2}$} on the data of \citet{venter} to
a fitted Gaussian probability density function with noise $\hnu = 0.19$. We estimate the noise level $\hnu = 0.19$
using a robust method that ignores outliers by Winsorizing the smallest and largest 5\% of the normalized differences \smash{$(Z_i-Z_i')/\surd{2}$} and rescales to maintain unbiasedness under Gaussian noise. The choice of Winsorization is motivated by the robustness of our method to underestimation of the noise level, as explained in Example~\ref{example:gaussian} and further demonstrated in the simulations of Section~\ref{subsec:misspecified_simulation} below.

The modeling choice \smash{$Z_i \mid U_i \sim \mathcal{N}(U_i, \,\hnu^2)$} illustrates our approach and serves as a starting point for analysis. We emphasize, however, that our method 
enables an epidemiologist
to posit a more realistic model of the noise density $p(z \mid u)$ based
on scientific understanding of CD4 counts and additional datasets
with repeated measurements.
Our causal identification strategy will be most credible when the scientist has considerable knowledge 
about the noise mechanism.
Henceforth in applying our approach, we assume that measurement error in the log CD4 counts follows
\smash{$Z_i \mid U_i \sim \mathcal{N}(U_i, \,\hnu^2)$}, where $U_i$ is the true underlying log CD4 count of patient $i$.
Given this noise model, we apply our noise-induced randomization (NIR) approach, with sensitivity model $\mathcal{T}_0$ to test for the existence of any treatment effects (as explained after Corollary~\ref{coro:valid_ci_het}).

\subsection{Method comparison and interpretation of results}
\label{subsec:art_method_comparison}
As a first comparison point, we consider treatment effect estimates obtained via the continuity-based
approach proposed by \citet{calonico2014robust}, which has recently become popular in applications.
This approach involves first fitting the regression discontinuity parameter
via local linear regression, and then estimating and correcting for its bias in a way that's
asymptotically justified under higher-order smoothness assumptions \citep{calonico2014robust}.
We implement this approach via the \texttt{R} package \texttt{rdrobust} of \citet{calonico2015rdrobust} with default tuning parameters.

As a second baseline, we  consider the minimax linear inference approach developed by
\citet{armstrong2018optimal,armstrong2016simple}, \citet{imbens2019optimized} and \citet{kolesar2018inference};
we use the \texttt{R} package \texttt{optrdd} of \citet{imbens2019optimized}.
This approach posits a constant $B$ such that \smash{$|\mu''_{(w)}(z)| \leq B$} for all
$w \in \cb{0, \, 1}$ and $z \in \RR$, and then provides intervals that are robust to the worst-case
bias under the curvature bound.
The main difficulty in using this approach is in choosing the curvature bound $B$.
We use the heuristic considered in \citet{armstrong2016simple}: We fit fourth-degree polynomials
to $\mu_{(0)}(\cdot)$ and $\mu_{(1)}(\cdot)$, and take the largest estimated curvature obtained anywhere.
Relative to rdrobust, the minimax linear inference approach makes explicit how smoothness is
used for inference (i.e., if one believes in the proposed curvature bound $B$, one should also believe in
the resulting intervals). In contrast, rdrobust relies more directly on asymptotics justified by
higher-order smoothness; see \citet{calonico2018effect} for further discussion.

\begin{table}
	\caption{Nominally 95\% confidence intervals for the effect of ART on retention rate of HIV patients,
  as given by our noise-induced randomization (NIR) method,
  rdrobust, and optrdd.   The curvature parameter for optrdd is chosen using the
  heuristic of \citet{armstrong2016simple}, resulting in $B=1.46$.}
\renewcommand{\arraystretch}{1.3}
{
  \begin{tabular}{lccc}
    \emph{Method} & NIR ($\mathcal{T}_0$) & rdrobust & optrdd \\ 
    \emph{95\% Confidence Interval} & 0.111 $\pm$ 0.102 & 0.170 $\pm$ 0.076 & 0.153 $\pm$ 0.080 \\  
  \end{tabular}
}
\label{tab:confidence_interval}
\end{table}

\begin{figure}
\centering 
\includegraphics[width=\textwidth]{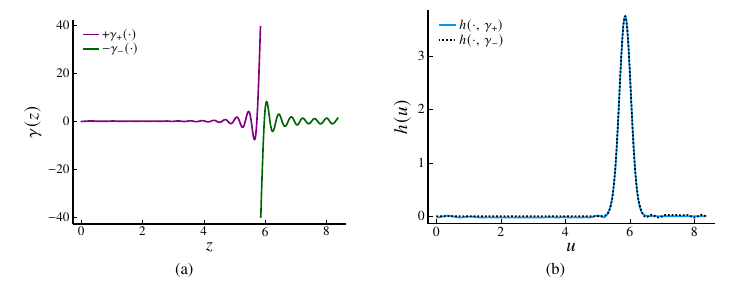}
\caption{Noise-induced randomization analysis of a regression discontinuity design with CD4 counts as the running variable. {(a)} $\gamma_\pm$ weighting functions vs. the running variable $z$.  {(b)} Implied latent weighting $h(u, \gamma_+)$, $h(u, \gamma_-)$ as a function of the latent $u$.}
\label{fig:bor_nts}
\end{figure}

We present the results in Table \ref{tab:confidence_interval}. All displayed confidence intervals
are significant at the 95\% level. What differs is the assumptions we need to justify
these confidence intervals. The baseline methods given here rely on quantifying the smoothness of the
$\mu_{(w)}(\cdot)$ in a data-driven way; and the credibility of the resulting intervals hinges on how well
we believe this task can be accomplished. In contrast, our NIR intervals are directly justified by the
posited measurement error model for the running variable $Z_i$ and the induced effective treatment randomization.

Whether practitioners prefer the NIR intervals or the continuity-based alternative will likely
depend on their intended use. Here, the continuity-based intervals are shorter than the NIR intervals, which
is desirable in settings where precision is at a premium. (In the simulation study, we show examples where the NIR intervals are shorter.)
On the other hand, the NIR intervals are explicitly justified using a form of
effective treatment randomization, and thus may be seen as getting closer to
best practices for credible causal inference as outlined by~\citet{rubin2008objective}. In some
settings, practitioners may want to report both:  One could see the NIR intervals as conservative intervals
that may sustain stricter scrutiny in terms of identification (by scrutiny of the assumed noise model), and the continuity-based ones as sharper
intervals if one is willing to rely on data-driven smoothness estimation.

Finally, for intuition, in Fig.~\ref{fig:bor_nts} we show the weighting functions $\gamma_\pm$ selected via
quadratic programming and that were used by the NIR approach (Section~\ref{sec:weighted_opt}),
and the implied latent weighting $h(\cdot, \gamma_+)$, $h(\cdot, \gamma_-)$ as per~\eqref{eq:hkernel_def}. Units with $Z_i$ close to the cutoff are strongly upweighted,
and so we achieve approximate balance in terms of the latent $U_i$.
The oscillations of the weighting functions $\gamma_{\pm}$ near the cutoff arise due to higher order bias corrections in nonparametric estimation and are common also for local linear regression estimates when represented as weighted averages (see, e.g,~\citealp[Fig. 1b]{gelman2019high}).

\subsection{Methodological detour: Noise-induced versus continuity-based inference?}
\label{sec:noise_to_continuity}

The two types of intervals discussed above may appear to rely on 
incomparable identification strategies. However, we can build a formal bridge connecting them.
One can verify, that in the presence of Gaussian measurement error, the functions $\mu_{(w)}(z) = \mathbb E\{Y_i(w) \mid Z_i=z\}$ must
be smooth. Under Assumptions \ref{assu:noise}--\ref{assu:exogenous},
$\mu_{(w)}(z) = \textstyle \int \alpha_{(w)}(u)p(z\mid u)\,dG(u) \,\big/\,\int p(z\mid u)\,dG(u)$,
so if $\alpha_{(w)}(\cdot)$ is bounded and $z \mapsto p(z \mid u)$ is continuous, then by the dominated convergence
theorem we can show that $\mu_{(w)}(\cdot)$ is also continuous \citep[e.g.,][Proposition 2]{lee2008randomized}. 
Furthermore, higher order differentiability of $p(\cdot \mid u)$ implies the same for $\mu_{(w)}(\cdot)$
\citep[e.g.,][Lemma A.1]{dong2021can}.
Here, we will investigate this connection to gain further insights on the relationship
between noise-induced-randomization and continuity-based methods shown above.

To this end, we define the worst-case possible curvature at $z$ among all data-generating distributions satisfying Assumptions~\ref{assu:noise}--\ref{assu:bounded} with conditional density $p(\cdot \mid \cdot)$ such that the marginal density of the running variable at $z$ is lower bounded by $\rho > 0$:
\begin{equation}
\label{eq:worst_case_curvature}
\Curv(z,\rho, p) = \sup\cb{\abs{\frac{d^2\mu_{(w)}(z)}{dz^2}}\,:\, f_G(z) = \textstyle \int p(z \mid u) dG(u) \geq \rho, \, \alpha_{(w)}(\cdot)\in [0,1]}.
\end{equation}
In~\eqref{eq:worst_case_curvature} we constrain ourselves to marginal densities such that $f_G(z) \geq \rho$ for $\rho>0$, because typically $\Curv(z,\,0, \, p) = \infty$. In Supplement~\ref{subsec:curvature_comp}, we explain how the quantity~\eqref{eq:worst_case_curvature} may be computed numerically for any sufficiently regular $p$.  One can then use the upper bounds on the second derivative of
\smash{$\mu_{(w)}(\cdot)$} in \eqref{eq:worst_case_curvature} in conjunction with, e.g., the estimators of \citet{imbens2019optimized}
and \citet{armstrong2016simple} that provide uniform inference for the regression discontinuity parameter
given a curvature bound on the response function.

To provide intuition for~\eqref{eq:worst_case_curvature}, we provide analytic lower and upper bounds on~\eqref{eq:worst_case_curvature} in the case of Gaussian measurement error, i.e., with
\smash{$Z_i \mid U_i \sim \nn(U_i, \, \nu^2)$} that quantify dependence on the noise level $\nu$ and the lower bound $\rho$ on the density. 
\begin{proposition}
\label{prop:continuity_bounds}
Suppose that Assumptions \ref{assu:noise}--\ref{assu:bounded} hold with noise model \smash{$Z_i \mid U_i \sim \nn\p{U_i, \, \nu^2}$}, where $\nu >0$. Then, \smash{$\mu_{(w)}(\cdot)$}
is infinitely differentiable and:
\begin{equation*}
\frac{-\log(2\pi \nu^2 \rho^2)}{10\nu^2}\leq \Curv\{z,\,\rho, \, \mathcal{N}(\cdot,\, \nu^2)\} \leq \frac{-18\log(\pi \nu^2 \rho^2)}{\nu^2}\;\text{ for all } z \in \RR,\, \rho \in \big(0,\,1 /\sqrt{2\pi \nu^2}\big).
\end{equation*}
\end{proposition}

We now return to the application of \citet{bor2017treatment}. Recall that we assumed a measurement error
with noise $\hnu = 0.19$. We estimate the density of the running variable at the cutoff as
$\hat{f}(c)= 0.57$ using the nonparametric maximum likelihood estimator. Using optrdd with
curvature parameter \smash{$\Curv\{c,\,\hat{f}(c), \, \mathcal{N}(\cdot,\, 0.19^2)\}$} (equal to $31.3$)
yields intervals that are directly justified by our noise model, just like noise-induced randomization. The resulting 95\% confidence interval is $\tau \in (0.071 \pm 0.130)$ which is
much wider than any of the intervals reported in Table \ref{tab:confidence_interval}.

The reason the optrdd intervals 
are wider than the continuity-based intervals in Table \ref{tab:confidence_interval} is
that our noise model implies much less continuity than is discovered by the data-driven
methods. For example, our noise model guarantees a curvature bound of $B = 31.3$, whereas the
heuristic of \citet{armstrong2016simple} gives
a curvature bound of $B = 1.46$ (i.e., it finds the function to be 20x smoother than guaranteed
by measurement error). This highlights the extent to which our proposal (and other methods justified by
measurement error alone) can be seen as stricter than continuity-based alternatives in terms of the
information used to estimate treatment effects.

\section{Application: Test Scores in Early Childhood}
\label{sec:education}

We next consider the behavior of our method in a semi-synthetic regression discontinuity design built using 
data from the Early Childhood Longitudinal Study~\citep{ECLS}. This dataset has
scaled mathematics test scores for $n = 18,174$ children from kindergarten to fifth grade.
Furthermore, each test score is accompanied by a noise variance obtained via
item response theory; see \citet{ECLS} for further details.

\begin{figure}
  \centering
  \includegraphics[width=\textwidth]{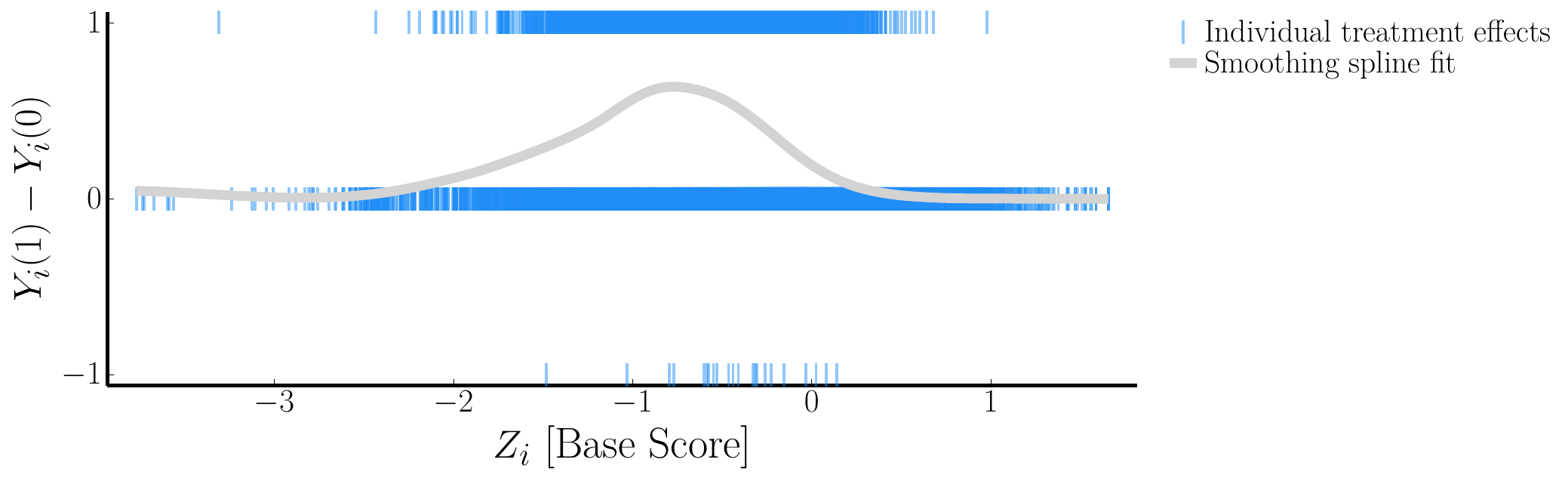}
  \caption{Scatterplot of the individual treatment effects $Y_i(1) - Y_i(0)$ against
  the running variable $Z_i$, with data derived from the Early Childhood Longitudinal
  Study~\citep{ECLS} as discussed in Section \ref{sec:education}. We fit the ground
  truth treatment effect function $\mathbb E\{Y_i(1) - Y_i(0) \mid Z_i = z\}$, shown as a line,
  using a smoothing spline.}
  \label{fig:ground_truth}
\end{figure}

Each sample $i = 1, \dotsc,  n$ is built using the sequence of test scores from a single child.
We set the running variable $Z_i$ to be the child's kindergarten spring semester score, and
set treatment as $W_i = \ind(Z_i \geq c)$ for a cutoff $c = -0.2$. We set
control potential outcomes $Y_i(0) \in \cb{0, \, 1}$ to indicate whether the child's score was
above $a = 0.5$ in spring semester of their first grade, while $Y_i(1) \in \cb{0, \, 1}$ measures the same
quantity in spring semester of their second grade; these are analogous to typically studied outcomes such as passing subsequent examinations. Thus, the treatment effect $Y_i(1) - Y_i(0)$
measures the child's improvement in passing the test (i.e., clearing the cutoff $a = 0.5$)
between first and second grades.

As shown in Fig.~\ref{fig:ground_truth}, there is considerable heterogeneity in the regression discontinuity
parameter $\tau_{c'} = \mathbb E\{Y_i(1) - Y_i(0) \mid Z_i = c'\}$ as we vary $c'$ away from the cutoff:
For children with either very good or very bad values of $Z_i$ the treatment effect is essentially
0 (since they will pass or, respectively, fail to pass the cutoff $a$ in both first and second grade
with high probability), while for students with intermediate values of $Z_i$ there is a large treatment
effect. We chose the parameters $a$ and $c$ in our semi-synthetic construction to
accentuate this type of heterogeneity.

\begin{figure}
  \centering
  \includegraphics[width=\textwidth]{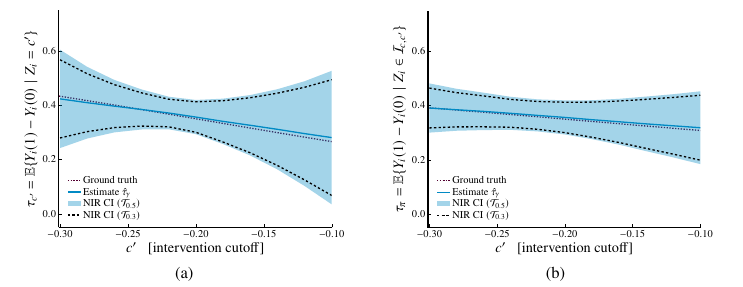}
  \caption{Regression discontinuity inference using our method (NIR) with data generated as in
  Fig.~\ref{fig:ground_truth}. {(a)} Estimates and  $95\%$ confidence intervals of the regression
  discontinuity parameter \eqref{eq:RDparam}, and {(b)} of the policy relevant parameter \eqref{eq:change_cutoff}.
  $\mathcal{I}_{c,c'}$ is the left-closed, right-open interval between $c$ and $c'$ when $c\neq c'$, and $\mathcal{I}_{c,c'} = \cb{c}$ for $c=c'=-0.2$.}
\label{fig:rddandpolicy}
\end{figure}

For this problem, the possibility and credibility of inference with noise-induced randomization builds on item response theory (IRT), a widely used model in educational testing. According to IRT, the $i$-th child's test score $Z_i$ is a noisy reflection of their true ability with (approximate) Gaussian measurement error with variance $\nu_i^2$ which is also determined by IRT. Below, we assume Gaussian errors in the running variable, i.e., \smash{$Z_i \mid U_i \sim \nn(U_i,\,\hnu^2)$},
and, following Example~\ref{example:gaussian}, we set $\hnu = \min_i \{\nu_i\} = 0.2043$ to match the lowest noise estimate provided
in the dataset. We run our method using sensitivity model $\mathcal{T}_{0.5}$. In this application, the monotonicity restriction $\tau(\cdot) \geq 0$ appears plausible, since the treatment effect measures the child's improvement
between first and second grades, and as explained after~\eqref{eq:sensi}, $\mathcal{T}_{0.5}$ does not place further restrictions on treatment effect heterogeneity. We also construct confidence intervals centered
at the same point estimates under $\mathcal{T}_{0.3}$; this sensitivity model is plausible based on the treatment heterogeneity in the ground truth individual treatment effects (Fig.~\ref{fig:ground_truth}).

Our main question is whether our procedure is able to estimate this heterogeneity, i.e.,
whether it can accurately recover variation in treatment effects away from the cutoff. To this
end, we consider two statistical targets: First, we consider estimation of the regression discontinuity
parameter \eqref{eq:RDparam} at $c'$ away from the cutoff, and second, the policy-relevant
parameter \eqref{eq:change_cutoff} quantifying the effect of changing the cutoff from $c$ to \smash{$c'$}. Results for both targets are shown in Fig.~\ref{fig:rddandpolicy}. Our method is able to recover heterogeneity. In both cases, the confidence intervals cover the ground truth.
They are narrowest near the cutoff $c = -0.2$, and get wider as we move away from the cutoff.

\section{Simulation Study}
\label{sec:simu}
\subsection{Well-specified discrete noise model}
\label{subsec:discrete}
To complement the picture given by our applications, we consider a simulation study to
assess the performance of our method in terms of its accuracy and coverage. We first consider a data-generating distribution with null treatment effects $\tau(u)=0$ wherein $Z_i$ has discrete support,
and has a binomial distribution conditionally on the latent $U_i$. For $i=1,\dotsc,n$, we generate,
\begin{align}
& U_i \sim \mathrm{Unif}(0.5, \, 0.9),\;\;\;\;  Z_i \mid U_i \sim \mathrm{Binomial}(K, U_i),\;\;\;\;  W_i = \ind(Z_i \geq 0.6K),  \label{eq:discrete_setting} \\ 
& Y_i(w) \mid U_i \sim \mathrm{Bernoulli}\{0.25\cdot\ind(u \leq c^*) \; + \; 0.75\cdot\ind(u > c^*)\}, \label{eq:simulation_response}
\end{align}
where the number of trials $K$ and number of samples $n$ are simulation parameters and $c^*=0.6$.

We compare the following point estimates and $95\%$ confidence intervals for the (null) treatment effect:
\begin{enumerate}
    \item Noise-induced randomization (NIR) with $p(\cdot \mid u) = \text{Binomial}(K, u)$ and using the sensitivity class $\mathcal{T}_0$ 
    (cf. justification after Corollary~\ref{coro:valid_ci_het}).
    \item Optrdd with a curvature upper bound $B$ as in~\eqref{eq:worst_case_curvature} applied for a binomial running variable (see Supplement~\ref{subsec:curvature_binomial} for details).
    \item Rdrobust as implemented  in the \texttt{R} package \texttt{rdrobust} of \citet{calonico2015rdrobust} with default specification and taking the debiased estimate as the point estimate.
\end{enumerate}

We evaluate methods by computing the confidence interval coverage, the expected half-length of confidence intervals and the mean absolute error. These metrics are computed by averaging over 1,000 Monte Carlo replications.

\begin{table}
\renewcommand{\arraystretch}{1.2}
\caption{Simulation results in the binomial noise setting in~\eqref{eq:discrete_setting},~\eqref{eq:simulation_response}. We vary the sample size $n$ and number of trials  $K$. We compare three methods (NIR, optrdd, rdrobust) and report the coverage of confidence intervals (coverage), the expected half-length of the intervals (length) and the mean absolute error (MAE). }
\centering
{\begin{tabular}{|r|r|c|cccccc|}
 \hline
 \multicolumn{2}{|c|}{} & $K$ & 5 & 10 &  25 & 50 &  100 & 200 \\ 
    \hline
  \multirow{9}{*}{\rotatebox[origin=c]{90}{$n = 1,000$}}    
& \multirow{3}{*}{\rotatebox[origin=c]{90}{coverage}}  
 & optrdd& 100.0\% & 99.8\% & 98.8\% & 98.0\% & 97.1\% & 96.5\% \\ 
 & & rdrobust & -- & -- & 94.2\% & 94.5\% & 93.4\% & 93.4\% \\ 
 & & NIR & 100.0\% & 97.1\% & 97.2\% & 97.6\% & 98.1\% & 98.6\% \\ 
 \cline{2-9}
 & \multirow{3}{*}{\rotatebox[origin=c]{90}{length}} 
 & optrdd & 0.452 & 0.383 & 0.347 & 0.344 & 0.370 & 0.423 \\ 
 & & rdrobust & -- & -- & 0.352 & 0.376 & 0.353 & 0.342\\ 
 & & NIR & 0.433 & 0.220 & 0.228 & 0.257 & 0.303 & 0.398 \\ 
    \cline{2-9}
  & \multirow{3}{*}{\rotatebox[origin=c]{90}{MAE}} 
  & optrdd& 0.089 & 0.095 & 0.113 & 0.119 & 0.130 & 0.153\\ 
 & & rdrobust &  -- & -- & 0.148 & 0.164 & 0.155 & 0.151 \\ 
 & & NIR &  0.068 & 0.076 & 0.084 & 0.091 & 0.105 & 0.126  \\ 
 \hline
  \multirow{9}{*}{\rotatebox[origin=c]{90}{$n = 2,000$}}    
& \multirow{3}{*}{\rotatebox[origin=c]{90}{coverage}}  
 & optrdd & 100.0\% & 100.0\% & 98.8\% & 98.5\% & 97.4\% & 96.6\%  \\ 
 & & rdrobust & -- & -- & 94.0\% & 94.0\% & 93.9\% & 92.8\% \\ 
 & & NIR & 100.0\% & 95.4\% & 95.9\% & 96.8\% & 97.1\% & 97.5\%  \\ 
 \cline{2-9}
 & \multirow{3}{*}{\rotatebox[origin=c]{90}{length}} 
 & optrdd & 0.396 & 0.325 & 0.280 & 0.273 & 0.287 & 0.322\\ 
 & & rdrobust & -- & -- & 0.244 & 0.261 & 0.246 & 0.239 \\ 
 & & NIR & 0.333 & 0.161 & 0.160 & 0.178 & 0.207 & 0.258\\ 
   \cline{2-9}
  & \multirow{3}{*}{\rotatebox[origin=c]{90}{MAE}} 
  & optrdd & 0.067 & 0.072 & 0.083 & 0.092 & 0.102 & 0.118 \\ 
 & & rdrobust & -- & -- & 0.105 & 0.117 & 0.110 & 0.106 \\ 
 & & NIR &0.052 & 0.063 & 0.061 & 0.065 & 0.076 & 0.093\\ 
 \hline
 \multirow{9}{*}{\rotatebox[origin=c]{90}{$n = 10,000$}}    
& \multirow{3}{*}{\rotatebox[origin=c]{90}{coverage}}  
 & optrdd &  100.0\% & 100.0\% & 100.0\% & 99.1\% & 98.3\% & 98.1\% \\ 
 & & rdrobust & -- & -- & 94.0\% & 94.6\% & 94.4\% & 94.4\% \\ 
 & & NIR &100.0\% & 96.4\% & 95.9\% & 96.3\% & 96.0\% & 96.4\% \\ 
 \cline{2-9}
 & \multirow{3}{*}{\rotatebox[origin=c]{90}{length}} 
&optrdd & 0.322 & 0.249 & 0.184 & 0.167 & 0.167 & 0.177 \\ 
 & & rdrobust & -- & -- & 0.104 & 0.115 & 0.107 & 0.108 \\ 
  && NIR & 0.220 & 0.078 & 0.074 & 0.081 & 0.093 & 0.111 \\ 
   \cline{2-9}
  & \multirow{3}{*}{\rotatebox[origin=c]{90}{MAE}} 
  &optrdd & 0.030 & 0.033 & 0.038 & 0.047 & 0.056 & 0.063 \\ 
  && rdrobust & -- & -- & 0.044 & 0.050 & 0.047 & 0.048 \\ 
  && NIR & 0.021 & 0.030 & 0.029 & 0.031 & 0.036 & 0.043 \\ 
 \hline
\end{tabular}}
\label{tab:simulation_binom_setup1}
\end{table}

The results of the simulation study are shown in Table~\ref{tab:simulation_binom_setup1}. All methods have approximately correct coverage, with optrdd and NIR always achieving the nominal $95\%$ level and rdrobust slightly undercovering. Although rdrobust and its distributional theory have been developed under the assumption of a continuous rather than discrete running variable, it performs reasonably well.
For small $K$ and $n$, rdrobust sometimes returns an error, in which case we do not report its performance.  NIR yields the shortest confidence intervals in most settings. The number of trials $K$ determines the effective noise level.
Our method performs best when there is more effective noise  in the running variable, that is, when $K$ is small ($K\leq 25$). This is in contrast to rdrobust, whose performance improves as $K$ increases and the running variable becomes less discrete, until at $K=200$ it leads to shorter confidence intervals than NIR. As expected, the confidence interval length decreases for all methods as the sample size $n$ increases.

This simulation experiment corroborates the claim that our method, NIR, can flexibly turn assumptions
about exogenous noise in the running variable $Z_i$ into a practical procedure for inference in regression discontinuity designs. We achieve nominal coverage
across simulation settings. 
Our results also point to the possibility that NIR may result in improved power in
settings where running variables are discrete with known noise. This would not be unreasonable,
as continuity-based approaches were not necessarily designed for this
setting (although, as discussed in \citet{kolesar2018inference} they can rigorously
be used given appropriate interpretation).

\subsection{Misspecified continuous noise model}
\label{subsec:misspecified_simulation}
We next explore the impact of misspecification of the noise model on the performance of NIR. We fix the sample size as $n=10,000$ and generate for $i=1,\dotsc,n$: $U_i \sim \nn(0, 1)$, $Z_i \mid U_i \sim p(\cdot \mid U_i)$, $W_i  = \ind(Z_i \geq 0)$. 
We consider the following three location-scale models for \smash{$p(\cdot \mid U_i)$}: Gaussian, 
t with 6 degrees of freedom, and Laplace.
In each case, the location is equal to $U_i$ and the scale is such that $\operatorname{Var}(Z_i \mid U_i) = 0.5^2$.
The response is generated as in~\eqref{eq:simulation_response} with $c^*=0$.

For each simulation setting,  we compare the following methods: 
rdrobust, and noise-induced randomization (NIR) with noise model $\nn(U_i, \nu^2)$ for $\nu \in \cb{0.3, 0.5, 0.7, 0.9}$ and the sensitivity class $\mathcal{T}_0$. NIR is well-specified only in one case: when it is applied with noise level $\nu=0.5$ and data is generated according to the Gaussian location-scale model.

\begin{figure}

\centering
\includegraphics[width=\textwidth]{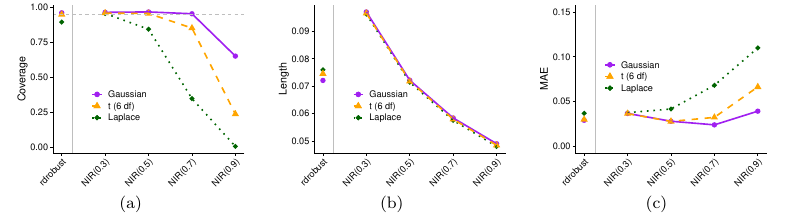}
\caption{
  Simulation with a continuous running variable. We compare rdrobust and NIR with working noise model $\nn(U_i, \nu^2)$ for $\nu \in \cb{0.3, 0.5, 0.7, 0.9}$ across three data generating processes wherein the true noise has variance $0.5^2$ and is Gaussian, t (with 6 degrees of freedom), and Laplace. NIR is well-specified only when $\nu=0.5$ and the noise model is Gaussian. We report (a) the coverage of confidence intervals, (b) the expected half-length of the confidence intervals, and (c) the mean absolute error.
}
\label{fig:misspecification}
\end{figure}

Our evaluation proceeds as in Section~\ref{subsec:discrete} and the results are shown in Fig.~\ref{fig:misspecification}. Rdrobust performs well across all three scenarios and sets a benchmark (even if it has some undercoverage with Laplace noise). With this standard in mind, we discuss the robustness of our proposed approach when confronted with a misspecified noise model. We begin by examining the situation where the true noise model is Gaussian. Then, NIR has the correct (95$\%$) coverage for $\nu \in \cb{0.3, 0.5, 0.7}$. Our theoretical results provide justification for $\nu=0.5$ (well-specification), and $\nu=0.3$ (underestimated noise-level as described in Example~\ref{example:gaussian}). Coverage for $\nu=0.7$ in the simulation is not justified theoretically, but demonstrates some robustness of NIR to the specification of the noise level. On the other hand, for $\nu=0.9$, the coverage of NIR drops to roughly 65$\%$, showing that NIR is not robust to substantial overestimation of the noise level. The expected half-length of NIR confidence intervals is decreasing in $\nu$. The mean absolute error exhibits a trade-off behavior, being minimized at $\nu = 0.7$, decreasing before this point, and then increasing thereafter. This suggests that for point estimation (rather than inference), $\nu$ acts similarly to a standard bias-variance trade-off parameter. NIR is also moderately robust to misspecification of the shape of the noise distribution: NIR with $\nu=0.3$ attains 95$\%$ coverage with Laplace noise, and NIR with $\nu \in \cb{0.3,0.5}$ attains nominal coverage with t-noise.  However, when both the noise level and the shape of the noise distribution are strongly misspecified, the coverage of NIR can be very low.

The results of this simulation study suggest that NIR is robust to moderate misspecification of the noise model. In applications, one should err toward underestimating the noise level if possible.

\section*{Software} All numerical results in this paper are reproducible with the code in the following Github repository: \url{https://github.com/nignatiadis/noise-induced-randomization-paper}.\\
We provide an implementation of NIR as a package in the Julia programming language~\citep{bezanson2017julia} that depends, among others, on JuMP.jl~\citep{DunningHuchetteLubin2017}.

\bibliographystyle{plainnat}
\bibliography{references}
\newpage
\appendix

\setcounter{page}{1}
\renewcommand{\thepage}{S\arabic{page}} 
\setcounter{equation}{0}
\renewcommand{\theequation}{S\arabic{equation}}
\setcounter{footnote}{0}
\renewcommand{\thefootnote}{S\arabic{footnote}}
\setcounter{figure}{0}
\renewcommand{\thefigure}{S\arabic{figure}}

\section{Additional figures}

\begin{figure}[h!]
\centering
\includegraphics{Tikz/dag_deconv.tikz}
\caption{Graphical illustration of the sharp regression discontinuity design
in which treatment is assigned  as a function of the latent variable $U$. $U$ is unobserved
and the analyst only observes $Z$, a noisy measurement of $U$. 
We contrast the above causal diagram with the causal diagram underlying noise-induced randomization
(Fig.~\ref{fig:dag}). In the latter case (noise-induced randomization),
 $Z$ is the running variable and the noise in $p(\cdot \mid \cdot)$
makes the inference task easier, that is, the regression discontinuity design
becomes more akin to a randomized
controlled trial. In the former case (not studied in this paper), the more noise there
is in $p(\cdot \mid \cdot)$, the more challenging the causal inference task becomes.
}
\label{fig:dag_deconv}
\end{figure}

\section{A further policy relevant estimand expressible as a weighted treatment effect }
\label{sec:measurement_error_estimand}

\begin{example}[Reducing measurement error]
Another policy intervention of potential interest could
involve switching to a more (or less) accurate device for measuring $Z_i$, thus changing the
noise level $\nu$ in the running variable. For example, one could imagine that a policy maker has the option
to reduce measurement error by using a new (potentially more expensive) measurement device,
and wants to know whether improved outcomes from more reproducible targeting are worth the cost.
Specifically, suppose that we currently assign treatment as $W_i = 1(Z_i \geq c)$ for
\smash{$Z_i \mid U_i \sim \nn(U_i, \, \nu^2)$}, and are considering a switch to a new treatment rule
$W_i' = 1(Z_i' \geq c)$ based on a measurement \smash{$Z_i' \mid U_i \sim \nn(U_i, \, {\nu'}^2)$}
with a different noise level $\nu'$. Writing $\Phi_\nu(\cdot)$ for the standard normal cumulative distribution
function with variance $\nu^2$ and assuming that $Z_i,Z_i'$ are independent conditionally on $U_i$, we see that the average treatment effect of patients who would be treated only with implementation of the policy change, is equal to
\begin{equation*}
\label{eq:reduce_measurement_error}
\mathbb E\{\tau(U_i) \mid W_i' > W_i\}
= \frac{\int \tau(u) \cb{1-\Phi_{\nu'}\p{c - u}}\Phi_{\nu}\p{c - u} dG(u)}{\int \cb{1-\Phi_{\nu'}\p{c - u}}\Phi_{\nu}\p{c - u} dG(u)},
\end{equation*}
which again is covered by \eqref{eq:specific_weighted_target}.
\end{example}

\section{Proofs}

\subsection{Proof of Proposition~\ref{prop:balance_response}}
\begin{proof}
Conditioning on the latent variable $U_i$, we find that
$$
\begin{aligned}
\EE{\gamma_+(Z_i)Y_i \cond U_i} &\stackrel{(i)}{=}  \EE{\gamma_+(Z_i) Y_i \cdot 1\p{Z_i\geq c} \cond U_i} \\
&\stackrel{(ii)}{=}  \EE{\gamma_+(Z_i) Y_i(1) \cdot 1\p{Z_i\geq c} \cond U_i}  \\
&\stackrel{(iii)}{=}\underbrace{\EE{Y_i(1) \cond U_i}}_{ \alpha_{(1)}(U_i)} \underbrace{\EE{\gamma_+(Z_i) 1\p{Z_i \geq c} \cond U_i}}_{h(U_i, \gamma_{+}) = \int \gamma_+(z) p(z\; | \;U_i) \, d\lambda(z)}
\end{aligned}
$$
In $(i)$ we used that $\gamma_+(z)=0$ for $z<c$, in $(ii)$ we used the fact that $Y_i = Y_i(1)$ for $Z_i \geq c$ by Assumption~\ref{assu:rdd} and in $(iii)$ we used exogeneity of the noise  (Assumption~\ref{assu:exogenous}). Finally, the expression for $\EE{\gamma_+(Z_i) 1\p{Z_i \geq c} \cond U_i} = \EE{\gamma_+(Z_i) \cond U_i}$ follows from Assumption~\ref{assu:noise}. By iterated expectation we thus find that $\EE{\gamma_+(Z_i)Y_i} = \EE{\alpha_{(1)}(U_i) h(U_i, \gamma_{+})}$. The proof for $\gamma_-$ is analogous.
\end{proof}

\subsection{Proof of Theorem~\ref{theo:consistency}}

\begin{proof}
By Assumption~\ref{assu:rdd}, the assumed moment conditions, and the law of large numbers, it follows that as $n\to \infty$, $\htau_\gamma - \taugamma \to 0$ in probability, where:
  \begin{equation}
  \label{eq:tau_gamma_general}
\taugamma = \mu_{\gamma,+}-\mu_{\gamma,-},\;\; \mu_{\gamma,+}=\frac{\mathbb E\{\gamma_+(Z_i)Y_i(1)\}}{\mathbb E\{\gamma_+(Z_i)\}},\;\; \mu_{\gamma,-} = \frac{\mathbb E\{\gamma_-(Z_i)Y_i(0)\}}{\mathbb E\{\gamma_-(Z_i)\}}.
\end{equation}
Under Assumptions~\ref{assu:noise} and~\ref{assu:exogenous}, 
the above expression is identical to~\eqref{eq:tau_gamma}. This follows from Proposition~\ref{prop:balance_response} and noting that $\mathbb E\{ \gamma_+(Z_i)\} = \mathbb E\{\hkernel{U_i}{\gamma_+}\}$, $\mathbb E\{ \gamma_-(Z_i)\} = \mathbb E\{\hkernel{U_i}{\gamma_-}\}$.
\end{proof}

\subsection{Proof of Corollary~\ref{coro:bias_decomposition}}
\begin{proof}
Noting that $\tau(u) = \alpha_{(1)}(u) - \alpha_{(0)}(u)$, this is proved by direct algebraic manipulation. 
\end{proof}

\subsection{Proof of Theorem~\ref{theo:main_clt}}

\begin{proof}
	\emph{Notation:} We write $\EE[n]{\cdot}$ to denote empirical averages, i.e., for a function $h(\cdot$), we write:
    $$ \EE[n]{h(Z_i)} = \frac{1}{n}\sum_{i=1}^n h(Z_i).$$
    We omit dependence on $n$ of the weighting kernels. We only prove a central limit theorem for $\hmu_{\gamma,+}$. The CLT for $\hmu_{\gamma,-}$ and $\htau_{\gamma} = \hmu_{\gamma,+} - \hmu_{\gamma,-}$ follow similarly.\\
	
	\noindent CLT for $\sum_i \gamma_+(Z_i)\p{Y_i(1)-\mu_{\gamma,+}}$: We seek to prove the following central limit theorem:
	
	$$\frac{\sum_{i=1}^n \gamma_+(Z_i)(Y_i(1) - \mu_{\gamma,+})}{\sqrt{n \EE{\gamma_+(Z_i)^2\p{Y_i(1)-\mu_{\gamma,+}}^2}}} \Rightarrow \nn\p{0,1}.$$
	We first note that the numerator has expectation $0$, since:
    $$\EE{\gamma_+(Z_i)(Y_i(1) - \mu_{\gamma,+})} =\EE{\gamma_+(Z_i) Y_i(1)} - \EE{\gamma_+(Z_i)}\frac{\EE{\gamma_+(Z_i)Y_i(1)}}{\EE{\gamma_+(Z_i)}} = 0.$$
    In the last step, we used the fact that as in~\eqref{eq:tau_gamma_general} and \eqref{eq:tau_gamma}
    \begin{equation}
    \label{eq:mugammaplus}
    \mu_{\gamma,+}=\frac{\EE{\alpha_{(1)}(U_i)\hkernel{U_i}{\gamma_+}}}{\EE{\hkernel{U_i}{\gamma_+}}} = \frac{ \EE{\gamma_+(Z_i)Y_i(1)}}{\EE{\gamma_+(Z_i)}},
    \end{equation}
    with equality of both numerators ($\mathbb E[\alpha_{(1)}(U_i)\hkernel{U_i}{\gamma_+}] = \mathbb E[\gamma_+(Z_i)Y_i(1)]$) and denominators ($\mathbb E[\hkernel{U_i}{\gamma_+}] = \mathbb E[\gamma_+(Z_i)]$).

  Next we will check the condition of Lyapunov's central limit theorem. Let $\ubar{\sigma}^2 = \inf_z \Var{Y_i \cond Z_i=z} > 0$. 
	\begin{equation}
    \label{eq:var_lower_bound}
	\begin{aligned}
	\Var{\gamma_+(Z_i)\p{Y_i(1)-\mu_{\gamma,+}}} &\geq \EE{\Var{\gamma_+(Z_i)\p{Y_i(1)-\mu_{\gamma,+}} \cond Z_i}} \\
	&= \EE{\gamma_+(Z_i)^2 \Var{Y_i(1)-\mu_{\gamma,+} \cond Z_i}} \\
	&= \EE{\gamma_+(Z_i)^2 \Var{Y_i(1) \cond Z_i}} \\
    &= \EE{\gamma_+(Z_i)^2 \Var{Y_i \cond Z_i}}\\
	&\geq  \ubar{\sigma}^2 \EE{\gamma_+(Z_i)^2}.
	\end{aligned}
	\end{equation}
	In the penultimate line we used the fact that $Y_i(1) = Y_i$ on $\cb{Z_i \geq c}$ and that $\gamma_+(z) = 0$ for $z<c$. We next bound $\mu_{\gamma,+}$ in~\eqref{eq:mugammaplus}. First, since $Y_i \in [0,1]$ by Assumption~\ref{assu:bounded}, it also follows that $\alpha_{(1)}(U) \in [0,1]$ almost surely. Thus:

  \begin{equation}
  \label{eq:mu_gamma_bounded}
    \abs{\mu_{\gamma,+}} =\abs{\frac{\EE{\alpha_{(1)}(U_i)\hkernel{U_i}{\gamma_+}}}{\EE{\gamma_+(Z_i)}}} \leq \frac{ \EE{\abs{\hkernel{U_i}{\gamma_+}}}}{\EE{\gamma_+(Z_i)} } \leq \frac{\sup_u \abs{\hkernel{u}{\gamma_+}}}{{\EE{\gamma_+(Z_i)} }} \leq C',
  \end{equation}
  for $n$ large enough. Then, for $q>0$ (and $n$ large enough) we have that:
	$$
	\begin{aligned}
	\EE{\abs{\gamma_+(Z_i)\p{Y_i(1)-\mu_{\gamma,+}}}^{2+q}} &\leq (C'+1)^{2+q}  \EE{\abs{\gamma_+(Z_i)}^{2+q}} \\
	&\leq (C'+1)^{2+q} \cdot \sup_z \abs{\gamma_+(z)}^q  \cdot \EE{\gamma_+(Z_i)^2}.
	\end{aligned}
	$$ 
	So:
	$$
	\begin{aligned}
	\frac{n\EE{\abs{\gamma_+(Z_i)\p{Y_i(1)-\mu_{\gamma,+}}}^{2+q}}}{\p{n \Var{\gamma_+(Z_i)\p{Y_i(1)-\mu_{\gamma,+}}}}^{(2+q)/2}} &\leq \frac{ (C'+1)^{2+q} \cdot \sup_z \abs{\gamma_+(z)}^q \cdot \EE{\gamma_+(Z_i)^2}}{n^{q/2} \cdot \ubar{\sigma}^{2+q} \cdot \EE{\gamma_+(Z_i)^2}^{(2+q)/2}} \\
	&\leq \p{\frac{C'+1}{\ubar{\sigma}}}^{2+q}\cdot \frac{\sup_z \abs{\gamma_+(z)}^q}{n^{q/2} \cdot \EE{\gamma_+(Z_i)^2}^{q/2}} \\
	&\leq \p{\frac{C'+1}{\ubar{\sigma}}}^{2+q}\cdot \frac{\sup_z \abs{\gamma_+(z)}^q}{n^{q/2}\EE{\gamma_+(Z_i)}^{q}} \\
	&\leq \p{\frac{C'+1}{\ubar{\sigma}}}^{2+q}\cdot \p{C n^{\beta - 1/2} }^q \to 0 \text{ as }n \to \infty.
	\end{aligned}
	$$
	This proves the central limit theorem.\\
	
	\noindent Estimation of normalization factor: Here we prove that $\EE[n]{\gamma_+(Z_i)}\big/ \EE{\gamma_+(Z_i)} = 1+o_{\mathbb P}(1)$. For any $\varepsilon >0$, by Chebyshev's inequality:
	$$
	\begin{aligned}
	\PP{ \abs{\EE[n]{\gamma_+(Z_i)} - \EE{\gamma_+(Z_i)}} \geq \varepsilon \EE{\gamma_+(Z_i)}} &\leq \frac{\Var{\gamma_+(Z_i)}}{n \varepsilon^2 \EE{\gamma_+(Z_i)}^2} \\
	&\leq \frac{\sup_z \gamma_+(z)^2}{n \varepsilon^2 \EE{\gamma_+(Z_i)}^2} \\
	& \leq \p{ \frac{C}{\varepsilon} \cdot n^{\beta-1/2}}^2 \to 0 \text{ as }n \to \infty.
	\end{aligned}
	$$
	
	\noindent CLT for $\hat{\mu}_{\gamma,+}$: Note that 
	$$\hat{\mu}_{\gamma,+} - \mu_{\gamma,+} = \sum_{i=1}^n \gamma_+(Z_i)(Y_i(1) - \mu_{\gamma,+})\big/ \sum_{i=1}^n \gamma_+(Z_i).$$
	The above display, along with our preceding result, $\mathbb E[\gamma_+(Z_i)] >0$, and Slutsky yield the CLT:

	$$\frac{\sqrt{n}\p{\hat{\mu}_{\gamma,+} - \mu_{\gamma,+} }}{\sqrt{\EE{\gamma_+(Z_i)^2\p{Y_i(1)-\mu_{\gamma,+}}^2}\big/\EE{\gamma_+(Z_i)}^2}}  \Rightarrow \nn\p{0,1}.$$
\end{proof}

\subsection{Proof of Proposition~\ref{prop:vgamma_consistent}}

\begin{proof}
	The proof here continues from the argument used for the proof of Theorem~\ref{theo:main_clt}. As we did there, we only prove the result for the variance of $\hmu_{\gamma,+}$, the result for $\htau_{\gamma}$ follows analogously. In the proof of Theorem~\ref{theo:main_clt} we already showed that $\EE[n]{\gamma_+(Z_i)}\big/ \EE{\gamma_+(Z_i)} = 1+o_{\mathbb P}(1)$. It thus suffices to show that:
    \begin{equation}
    \label{eq:variance_estimator_numerator}
    \EE[n]{\gamma_+(Z_i)^2\p{Y_i(1)-\hmu_{\gamma,+}}^2}\big /\EE{\gamma_+(Z_i)^2\p{Y_i(1)-\mu_{\gamma,+}}^2} = 1 +o_{\mathbb P}(1).
    \end{equation}
    We start by arguing that: 
	\begin{equation}
    \label{eq:oracle_variance_estimator}
    \EE[n]{\gamma_+(Z_i)^2\p{Y_i(1)-\mu_{\gamma,+}}^2}\big/\EE{\gamma_+(Z_i)^2\p{Y_i(1)-\mu_{\gamma,+}}^2} = 1 +o_{\mathbb P}(1).
    \end{equation}
    First:
    $$
    \begin{aligned}
    \Var{\frac{\EE[n]{\gamma_+(Z_i)^2\p{Y_i(1)-\mu_{\gamma,+}}^2}}{\EE{\gamma_+(Z_i)^2\p{Y_i(1)-\mu_{\gamma,+}}^2}}} &= \frac{\Var{\gamma_+(Z_i)^2\p{Y_i(1)-\mu_{\gamma,+}}^2}}{n \cdot \EE{\gamma_+(Z_i)^2\p{Y_i(1)-\mu_{\gamma,+}}^2}^2} \\
    &\leq \frac{ \EE{ \cb{\gamma_+(Z_i)^2\p{Y_i(1)-\mu_{\gamma,+}}^2} \cdot \cb{\gamma_+(Z_i)^2\p{Y_i(1)-\mu_{\gamma,+}}^2}}}{n \cdot \EE{\gamma_+(Z_i)^2\p{Y_i(1)-\mu_{\gamma,+}}^2}^2 }  \\
    & \leq \frac{  (C'+1)^2 \sup_{z} \abs{\gamma_+(z)}^2 }{n \EE{\gamma_+(Z_i)^2\p{Y_i(1)-\mu_{\gamma,+}}^2}} \to 0 \text{ as } n \to \infty.
    \end{aligned}
	$$
    Note that we verified that the last expression converges to $0$ as $n \to \infty$ during the verification of Lyapunov's condition in the proof of Theorem~\ref{theo:main_clt}. 
    It follows that the asymptotic convergence in~\eqref{eq:oracle_variance_estimator}
    holds in $L^2$, thus also in probability. 
    It remains to show that the feasible estimator in~\eqref{eq:variance_estimator_numerator} is asymptotically equivalent. We have the decomposition:
	$$
	\begin{aligned}
	&\gamma_+(Z_i)^2\p{Y_i(1)-\hat{\mu}_{\gamma,+}}^2 - \gamma_+(Z_i)^2\p{Y_i(1)-\mu_{\gamma,+}}^2 \\
	=\;& \gamma_+(Z_i)^2\p{\hat{\mu}_{\gamma,+}-\mu_{\gamma,+}}^2 + 2\gamma_+(Z_i)^2\p{Y_i(1)-\mu_{\gamma,+}}\p{\mu_{\gamma,+}-\hat{\mu}_{\gamma,+}}.
	\end{aligned}
	$$
	From the CLT of Theorem~\ref{theo:main_clt}, we know that:
	$$\p{\hat{\mu}_{\gamma,+} - \mu_{\gamma,+}}^2 = O_{\mathbb P}\p{n^{-1} \EE{\gamma_+(Z_i)^2\p{Y_i(1)-\mu_{\gamma,+}}^2}\bigg/\EE{\gamma_+(Z_i)}^2} = O_{\mathbb P}\p{n^{-1 + 2\beta}} = o_{\mathbb P}(1),$$
	and so:
	$$ \frac{\EE[n]{\gamma_+(Z_i)^2}}{\EE{\gamma_+(Z_i)^2\p{Y_i(1)-\mu_{\gamma,+}}^2}}\cdot\p{\hat{\mu}_{\gamma,+}-\mu_{\gamma,+}}^2 = O_{\mathbb P}(1) \cdot o_{\mathbb P}(1) = o_{\mathbb P}(1).$$
	The fact that the first term is $O_{\mathbb P}(1)$ follows by arguing with Chebyshev's inequality. First note that from~\eqref{eq:var_lower_bound}, we know that \smash{$\mathbb E[\gamma_+(Z_i)^2\p{Y_i(1)-\mu_{\gamma,+}}^2] \geq \ubar{\sigma}^2 \mathbb E[\gamma_+(Z_i)^2]$} and so it suffices to show that $\EE[n]{\gamma_+(Z_i)^2}/\EE{\gamma_+(Z_i)^2}$ is $O_{\mathbb P}(1)$. Indeed this term is $1+o_{\mathbb P}(1)$, since for any $\varepsilon > 0$:
    $$
    \begin{aligned}
    \PP{ \abs{\EE[n]{\gamma_+(Z_i)^2} - \EE{\gamma_+(Z_i)^2}} \geq \varepsilon \EE{\gamma_+(Z_i)^2}} &\leq \frac{\Var{\gamma_+(Z_i)^2}}{n \varepsilon^2 \EE{\gamma_+(Z_i)^2}^2} \\
    &\leq \frac{\sup_z \gamma_+(z)^2}{n \varepsilon^2 \EE{\gamma_+(Z_i)}^2} \cdot \frac{ \EE{\gamma_+(Z_i)^2}}{ \EE{\gamma_+(Z_i)^2}}  \\
    & \leq \p{ \frac{C}{\varepsilon} \cdot n^{\beta-1/2}}^2 \to 0 \text{ as }n \to \infty.
    \end{aligned}
    $$
    This proves the first term is negligible. To show that the second term is negligible, our basic argument is that
   $$ \frac{\EE[n]{\gamma_+(Z_i)^2\p{Y_i(1)-\mu_{\gamma,+}}}}{\EE{\gamma_+(Z_i)^2\p{Y_i(1)-\mu_{\gamma,+}}^2}}\cdot\p{\hat{\mu}_{\gamma,+}-\mu_{\gamma,+}} = O_{\mathbb P}(1) \cdot o_{\mathbb P}(1) = o_{\mathbb P}(1),$$
   and it remains to prove that the first term is indeed $O_{\mathbb P}(1)$. By Cauchy-Schwarz
   $$
   \begin{aligned}
   \abs{\EE[n]{\gamma_+(Z_i)^2\p{Y_i(1)-\mu_{\gamma,+}}}} &= \abs{\EE[n]{\gamma_+(Z_i)\cdot \gamma_+(Z_i)\p{Y_i(1)-\mu_{\gamma,+}}}} \\
   &\leq \p{ \EE[n]{\gamma_+(Z_i)^2}}^{1/2} \p{\EE[n]{ \gamma_+(Z_i)^2\p{Y_i(1)-\mu_{\gamma,+}}^2}}^{1/2}
   \end{aligned}
   $$
   But the above is the product of two \smash{$O_{\mathbb P}(\mathbb E[\gamma_+(Z_i)^2(Y_i(1)-\mu_{\gamma,+})^2]^{1/2})$} terms (as we showed above), so we conclude upon dividing by \smash{$\mathbb E[\gamma_+(Z_i)^2(Y_i(1)-\mu_{\gamma,+})^2]$}.
\end{proof}

\subsection{Proof of Proposition~\ref{prop:validbiasbound}}

\begin{proof}
Consider the event $\cb{G \in \mathcal{G}_n}$. On this event, by definition we have $\abs{b_{\gamma}} \leq \hB_{\gamma,M}$.
This implies that 
$$\cb{G \in \mathcal{G}_n} \subset \cb{\abs{b_{\gamma}} \leq \hB_{\gamma, M}},$$
and so
$$\PP{\abs{b_{\gamma}} \leq \hB_{\gamma, M}} \geq \PP{G \in \mathcal{G}_n}.$$
It thus suffices to show that the RHS converges to $1$ as $n \to \infty$. By construction of $\mathcal{G}_n$ in~\eqref{eq:calG} and Massart's tight constant for the DKW inequality~\citep{massart1990tight}, it holds that
$$\PP{G \in \mathcal{G}_n} \geq \PP{  \sup_{t \in \mathbb R}\abs{F(t) - \widehat{F}_n(t)}  \leq  \sqrt{\log\p{2/\alpha_n}\big/(2n)}} \geq 1-\alpha_n.$$
Since $\alpha_n \to 0$, we conclude.
\end{proof}

\subsection{Proof of Corollary~\ref{coro:valid_ci}}
By Theorem~\ref{theo:main_clt}, Proposition~\ref{prop:vgamma_consistent}, and Slutsky, we have that
\begin{equation}
\label{eq:suppl_clt}
\frac{\sqrt{n}\p{\hat{\tau}_{\gamma} - \tau_w - b_{\gamma} }}{\hV_{\gamma}^{1/2}}  \Rightarrow \nn\p{0,1},
\end{equation}
where $b_{\gamma} = \taugamma-\tau_w$ by definition. We make the dependence of $\ell_{\alpha}$ (defined in~\eqref{eq:CI}) on $\hB_{\gamma,M}, \hV_\gamma/ n$ explicit by writing $\ell_{\alpha}(\hB_{\gamma,M}, \hV_\gamma/ n)$. Write $A_n = \cb{\abs{b_{\gamma}} \leq \hB_{\gamma, M}}$. Then, on the event $A_n$, it holds that $\ell_{\alpha}(\hB_{\gamma,M}, \hV_\gamma/ n) \geq \ell_{\alpha}(b_{\gamma}, \hV_\gamma/ n)$. Thus:
$$
\begin{aligned}
\PP{ \tau_w \in \hat{\tau}_{\gamma} \pm \ell_{\alpha}(\hB_{\gamma,M}, \hV_\gamma/ n)} &\geq \PP{ \cb{\tau_w \in \hat{\tau}_{\gamma} \pm \ell_{\alpha}(b_{\gamma}, \hV_\gamma/ n)} \cap A_n} \\ 
&= \PP{\tau_w \in \hat{\tau}_{\gamma} \pm \ell_{\alpha}(b_{\gamma}, \hV_\gamma/ n)} + o(1), 
\end{aligned}
$$
where the last step holds since $\PP{A_n} \to 1$ by Proposition~\ref{prop:validbiasbound}. To proceed, we require some properties of $\ell_{\alpha}$ which is studied, e.g., in~\citet{armstrong2018optimal}, also see Figure 1a in~\citet{ignatiadis2022rejoinder} for a graphical representation. For any $b \in \RR$, $\sigma >0$, it holds that $\ell_{\alpha}(b, \sigma^2) = \sigma \ell_{\alpha}(b/\sigma, 1)$.
Furthermore $\ell_{\alpha}(b, 1) \geq q_{\alpha/2}$ for all $b$ and $\ell_{\alpha}(b, 1) - b \to q_{\alpha}$ as $b \to \infty$, where $q_{\alpha}$ is the $1-\alpha$ standard normal quantile. The function $b \mapsto \ell_{\alpha}(b,1)$ is uniformly continuous in $b$.

To simplify notation, let us also write $\hat{\sigma} = n^{-1/2} \hV_{\gamma}^{1/2}$, $\sigma = n^{-1/2}V_{\gamma}^{1/2}$. We find that:
$$
\begin{aligned}
&\PP{\tau_w \in \hat{\tau}_{\gamma} \pm \ell_{\alpha}(b_{\gamma}, \hV_\gamma/ n)}  \\ 
\;\;\;\;& = \PP{-\hat{\sigma}^{-1}\cb{\ell_{\alpha}(b_{\gamma}, \hat{\sigma}^2) + b_{\gamma}}\leq \hat{\sigma}^{-1}\p{\hat{\tau}_{\gamma} - \tau_w - b_{\gamma}} \leq \hat{\sigma}^{-1}\cb{\ell_{\alpha}(b_{\gamma}, \hat{\sigma}^2) - b_{\gamma}}} \\ 
\;\;\;\;& = \PP{-\ell_{\alpha}(b_{\gamma}/\hat{\sigma}, 1) - b_{\gamma}/\hat{\sigma} \leq \p{\hat{\tau}_{\gamma} - \tau_w - b_{\gamma}}/\hat{\sigma} \leq  {\ell_{\alpha}(b_{\gamma}/\hat{\sigma}, 1) - b_{\gamma}/\hat{\sigma}}}.
\end{aligned}
$$
To continue, we study $\ell_{\alpha}(b_{\gamma}/\hat{\sigma}, 1) - b_{\gamma}/\hat{\sigma}$ and $-\ell_{\alpha}(b_{\gamma}/\hat{\sigma}, 1)- b_{\gamma}/\hat{\sigma}$. It is important to recall the triangular array asymptotics in which $b_{\gamma}, \sigma, \hat{\sigma}_n$ depend on $n$, i.e., $b_{\gamma}=b_{\gamma^{(n)}}$, $\sigma = \sigma_n$, $\hat{\sigma} = \hat{\sigma}_n$.

We assume henceforth that $b_{\gamma^{(n)}} \geq 0$ for sufficiently large $n$ (otherwise we may consider limits along the subsequences $\{n: b_{\gamma^{(n)}} \geq 0\}$, and $\{n: b_{\gamma^{(n)}} < 0\}$). 

Next, let $0 < a_n \to \infty$ be a deterministic sequence such that
$$ 
\abs{\frac{\hat{\sigma}_n}{\sigma_n} - 1} = o_{\mathbb P}(a_n^{-1}), \;\; \abs{\frac{\sigma_n}{\hat{\sigma}_n} - 1} = o_{\mathbb P}(a_n^{-1}).
$$
Such a sequence exists by Proposition~\ref{prop:vgamma_consistent}.
Also note that $b_{\gamma^{(n)}} = O(1)$ since $\tau_w \in [-1,1]$ and $\theta_{\gamma}$ may be bounded as in~\eqref{eq:mu_gamma_bounded}.

We now partition $\mathbb N$ into $\mathcal{S}_{\geq} = \{n : b_{\gamma^{(n)}}/\sigma_n \geq \alpha_n\}$ and $\mathcal{S}_{<} =  \{n : b_{\gamma^{(n)}}/\sigma_n < \alpha_n\}$. We assume that the cardinality of both $\mathcal{S}_{<}$, $\mathcal{S}_{\geq}$ is infinite and study limits along these two subsequences (if the cardinality of either $\mathcal{S}_{<}$ or $\mathcal{S}_{\geq}$ is finite, then it suffices to study limits along the infinite cardinality one).

\noindent Limits along $\mathcal{S}_{<}$: In this case, we have that:
$$ \frac{b_{\gamma^{(n)}}}{\sigma_n} - \frac{b_{\gamma^{(n)}}}{\hat{\sigma}_n} = \frac{b_{\gamma^{(n)}}}{\sigma_n}\p{ 1 - \frac{\sigma_n}{\hat{\sigma}_n}} = o_{\mathbb P}(1).$$
Hence, by uniform continuity of $\ell_{\alpha}(\cdot, 1)$, we find that:
$$\ell_{\alpha}(b_{\gamma^{(n)}}/\sigma_n, 1) -  \ell_{\alpha}(b_{\gamma^{(n)}}/\hat{\sigma}_n, 1) = o_{\mathbb P}(1).$$
The central limit theorem in~\eqref{eq:suppl_clt} implies that the distribution function of the (asymptotic) pivot converges to the standard normal distribution function $\Phi(\cdot)$ uniformly. Furthermore $\Phi(\cdot)$ is uniformly continuous. So, letting $\tilde{Z} \sim \nn\p{0,1}$ independent of everything else, we get:
$$
\begin{aligned}
&\PP{-\ell_{\alpha}(b_{\gamma}/\hat{\sigma}, 1) - b_{\gamma}/\hat{\sigma} \leq \p{\hat{\tau}_{\gamma} - \tau_w - b_{\gamma}}/\hat{\sigma} \leq  {\ell_{\alpha}(b_{\gamma}/\hat{\sigma}, 1) - b_{\gamma}/\hat{\sigma}}} \\ 
&\;\;\;\;=\PP{-\ell_{\alpha}(b_{\gamma}/\sigma, 1) - b_{\gamma}/\sigma \leq \tilde{Z} \leq  {\ell_{\alpha}(b_{\gamma}/\sigma, 1) - b_{\gamma}/\sigma}} + o(1) \\ 
&\;\;\;\; \geq 1-\alpha + o(1),
\end{aligned}
$$
where in the last line we used the definition of $\ell_{\alpha}(b, 1)$. \\

\noindent Limits along $\mathcal{S}_{\geq}$: In this case, we have the following results. First,  $-\ell_{\alpha}(b_{\gamma}/\hat{\sigma}, 1)- b_{\gamma}/\hat{\sigma} \to -\infty$ in probability. Second, $\ell_{\alpha}(b_{\gamma}/\hat{\sigma}, 1) - b_{\gamma}/\hat{\sigma} \to q_{\alpha}$ in probability. Hence:
$$
\begin{aligned}
&\PP{-\ell_{\alpha}(b_{\gamma}/\hat{\sigma}, 1) - b_{\gamma}/\hat{\sigma} \leq \p{\hat{\tau}_{\gamma} - \tau_w - b_{\gamma}}/\hat{\sigma} \leq  {\ell_{\alpha}(b_{\gamma}/\hat{\sigma}, 1) - b_{\gamma}/\hat{\sigma}}} \\ 
&\;\;\;\;=\PP{\tilde{Z} \leq  q_{\alpha}} + o(1) \\ 
&\;\;\;\; = 1-\alpha + o(1).
\end{aligned}
$$
We conclude by combining all results above.

\subsection{Proof of Corollary~\ref{coro:valid_ci_het}}
Let $\tau_{h,+}$ be the convenience-weighted treatment effect defined in the statement of the corollary. Arguing as in~\eqref{eq:mu_gamma_bounded} we can show that $\tau_{h,+}$ is bounded (uniformly in $n$).

We will show that 
$$\abs{\operatorname{Bias}\{\gamma_{\pm},\, \tau_{h,+}; \, \alpha_{(0)}(\cdot),\, \tau(\cdot),\,G\}} \leq \hB_{\gamma,M'},$$
where $\hB_{\gamma,M'}$ is defined as in~\eqref{eq:biasfractional} for the estimand $\tau_w$ (rather than $\tau_{h,+}$). We make this dependence explicit by writing $\hB_{\gamma,M'} = \hB_{\gamma,M',\tau_w}$. The heterogeneity bias in Corollary~\ref{coro:bias_decomposition} vanishes for $\tau_{h,+}$, i.e.,
$$\operatorname{Bias}\{\gamma_{\pm},\, \tau_{h,+}; \, \alpha_{(0)}(\cdot),\, \tau(\cdot),\,G\} = \int\p{\frac{\hkernel{u}{\gamma_+}}{\EE[G]{\hkernel{U_i}{\gamma_+}}} - \frac{\hkernel{u}{\gamma_-}}{\EE[G]{\hkernel{U_i}{\gamma_-}}}} \alpha_{(0)}(u)\, dG(u).$$
However, the heterogeneity bias also vanishes when $\tau(\cdot) \in \mathcal{T}_0$ (for any choice of estimand $\tau_w$), so that:
$$ \abs{\operatorname{Bias}\{\gamma_{\pm},\, \tau_{h,+}; \, \alpha_{(0)}(\cdot),\, \tau(\cdot),\,G\}} = \abs{\operatorname{Bias}\{\gamma_{\pm},\, \tau_w; \, \alpha_{(0)}(\cdot),\, \tau(\cdot),\,G\}}, \text{ when } \tau(\cdot) \in \mathcal{T}_0.$$
Taking the supremum on the RHS over $\alpha_{(0)}(\cdot) \in [0,1]$, $\tau(\cdot) \in \mathcal{T}_{0}$ and $G \in \mathcal{G}_n$ demonstrates that:
$$\abs{\operatorname{Bias}\{\gamma_{\pm},\, \tau_{h,+}; \, \alpha_{(0)}(\cdot),\, \tau(\cdot),\,G\}} \leq \hB_{\gamma,0,\tau_w} \leq \hB_{\gamma,M',\tau_w}.$$
The conclusion follows as in the proofs of Proposition~\ref{prop:validbiasbound} and Corollary~\ref{coro:valid_ci}.

\subsection{Proof of Proposition~\ref{prop:suff_weights}}

\begin{proof}
    We provide the argument for $\gamma_+=\gamma_+^{(n)}$; the argument for $\gamma_-=\gamma_-^{(n)}$ is analogous. First note that \smash{$\sup_{z} \abs{\gamma_{+}(z)} > 0$} must hold, otherwise constraint~\eqref{eq:integrate_to_1} of the optimization problem would not be satisfied. For convenience we define the events:
    $$ B_{n,1} = \cb{ 1/k < \bar{F}([c,\infty))< 1-1/k,\; \sup_u \abs{\bar{w}(u)} < k},\; B_{n,2} = \cb{ \int_{[c,\infty)} \gamma_+(z) dF_G(z) > \delta}.$$
    Next, let $C$ be as in~\eqref{eq:optimization_clt} and $\tilde{C} = C/\delta$. Then, on the event $B_{n,2}$ we have that:
    $$\sup_{z} \abs{\gamma_{+}(z)} \leq C n^{\beta} \leq \tilde{C} n^{\beta} \cdot \delta < \tilde{C} n^{\beta}  \int_{[c,\infty)} \gamma_+(z) dF_G(z).$$
    Next, we will bound $\abs{\hkernel{u}{\gamma_+}}$ on the event $B_{n,1}\cap B_{n,2}$.
    Consider the weighting kernel $\tilde{\gamma}_+(z) = \ind(z \geq c) / \bar{F}([c,\infty))$ and $\tilde{\gamma}_-(z) = \ind(z < c)/(1-\bar{F}([c,\infty)))$. This is a feasible solution under the constraints of optimization problem~\eqref{eq:qp}, since $\sup_u \abs{\tilde{\gamma}_{\diamond}(u)} \leq k \leq Cn^{\beta}$, $\diamond \in \cb{+,-}$  (on the event $B_{n,1}\cap B_{n,2}$ and for $n$ large enough), $\int \tilde{\gamma}_-(z)d\bar{F}(z) = 1$ and $\int \tilde{\gamma}_+(z)d\bar{F}(z) = 1$. We upper bound the objective of the optimization problem for that choice of weighting kernel. First, the variance-proxy term in the objective is equal to $\sqb{\bar{F}([c,\infty))^{-1} + (1-\bar{F}([c,\infty)))^{-1}}/n$ which is $\leq 2k/n \leq 1$ for $n$ large enough. We also have that:
    $$\abs{h(u, \tilde{\gamma}_+)} = \abs{\bar{F}([c,\infty))^{-1} \cdot \int_{[c,\infty)}p(z \mid u)d\lambda(z)} \leq \bar{F}([c,\infty))^{-1} \leq k,$$
    and similarly $\abs{h(u, \tilde{\gamma}_-)} \leq k$. Hence by the triangle inequality we may bound the ``$t_1$'' term of the variance objective by $2k$, and similarly for the ``$t_2$'' term (recall that $\abs{\bar{w}(u)} \leq k$ on $B_{n,1}$ and that $0<M \leq 1$). Thus the objective of the whole optimization problem is upper bounded by $1+16k^2$. Thus, the objective for any optimal $\gamma_{\pm}$ of~\eqref{eq:qp} must be $\leq 1+16k^2$, which implies that
    $$ M \abs{ h(u, \gamma_+) - \bar{w}(u)} \leq \sqrt{1 + 16k^2} \leq 5k,$$
    and so
    $$ \sup_u \abs{h(u, \gamma_+)} \leq \sup_u \cb{ \abs{h(u, \gamma_+) - \bar{w}(u)} + \abs{\bar{w}(u)}} \leq 5k/M + k \leq 6k/M.$$
    We conclude that for $\tilde{C}$ as above and $\tilde{C}' = 6k/(M\delta)$, it holds that:
$$
\begin{aligned}
&\PP{\sup_{z} \abs{\gamma_{+}^{(n)}(z)} < \tilde{C} n^{\beta} \EE{\gamma_{+}^{(n)}(Z_i)},\;\; \sup_u \abs{\hkernel{u}{\gamma_{+}^{(n)}}} < \tilde{C}' \EE{\gamma_{+}^{(n)}(Z_i)}}\\
\geq \;\; &\PP{B_{n,1}\cap B_{n,2}} \to 1 \text{ as } n \to \infty.
\end{aligned}
$$
\end{proof}

\subsection{Proof of Proposition~\ref{prop:continuity_bounds}}

\begin{proof}
For the first result, note that $\mu_{(w)}(z)$ may in fact be extended to an analytic function across all of $\mathbb C$, cf.~\cite{kim2014minimax}. We proceed with the quantitative claims and first note that it suffices to consider the standard normal case, i.e., $\nu=1$. To see this, take \smash{$Z_i \mid U_i \sim \nn\p{U_i, \nu^2}$}. Then \smash{$\tilde{Z}_i = Z_i/\nu \mid U_i \sim \nn\p{U_i/\nu, 1}$} and we may apply the results to $\tilde{Z}_i$. Concretely, let \smash{$\tilde{m}:\RR \mapsto \RR$} be an arbitrary function and $m: z \mapsto \tilde{m}(z/\nu) = \tilde{m}(\tilde{z})$. This defines a bijection between functions that enables us to translate results for $\tilde{Z}_i$ into results for $Z_i$ and vice versa (by applying the chain rule). It only remains to express the density \smash{$\tilde{f}(\tilde{z})$} of $\tilde{Z}_i$ at $\tilde{z} = z/\nu$ in terms of the density $f$ of $Z_i$; by transformation we have $\tilde{f}(\tilde{z}) = \nu \cdot f(z)$. Furthermore, we derive all of our results for $\mu_{(0)}(z)$; the arguments for $\mu_{(1)}(z)$ are identical.\\

\noindent Upper bound: Fix $\tilde{c}>0$. Let $\tilde{\alpha}_{(0)}(u) = \tilde{c} + \alpha_{(0)}(u) \in [\tilde{c}, 1+\tilde{c}]$. Let $H \ll G$ be the probability measure with 
$$\frac{dH}{dG}(u) = \frac{\tilde{\alpha}_{(0)}(u)}{\int \tilde{\alpha}_{(0)}(u)dG(u)},$$
and write $h(z) = \int \varphi(z-u) dH(u)$. Then:
$$\mu_{(0)}(z) = \EE{ \alpha_{(0)}(U_i) \cond Z_i=z} =  \EE{ \tilde{\alpha}_{(0)}(U_i) \cond Z_i=z} - \tilde{c} =  \frac{h(z) \cdot \int \tilde{\alpha}_{(0)}(u)dG(u)}{f(z)} - \tilde{c}.$$
Taking the derivative with respect to $z$ (an operation which we denote interchangeably by $d/dz$ and $'$):
$$ \frac{d}{dz}\mu_{(0)}(z) = \int \tilde{\alpha}_{(0)}(u)dG(u) \cdot \p{ \frac{h'(z)}{f(z)} - \frac{h(z)}{f(z)} \cdot \frac{f'(z)}{f(z)}} = \int \tilde{\alpha}_{(0)}(u)dG(u) \cdot \frac{h(z)}{f(z)} \cdot \p{ \frac{h'(z)}{h(z)} -  \frac{f'(z)}{f(z)}}.$$
We next bound the three terms appearing in the expression above. First, we already saw that $\int \tilde{\alpha}_{(0)}(u)dG(u) \cdot \frac{h(z)}{f(z)} = \mu_{(0)}(z) + \tilde{c}$ with $\mu_{(0)}(z) \in [0,1]$ and so this term is upper bounded in absolute value by $1+\tilde{c}$.  Next, by Lemma A.1. in~\citet{jiang2009general} (which we state and prove at the end of this section for self-containedness) it holds that:
$$\abs{\frac{f'(z)}{f(z)}} \leq  \sqrt{-\log(2\pi f^2(z))},\;\; \abs{\frac{h'(z)}{h(z)}} \leq  \sqrt{-\log(2\pi h^2(z))}.$$
It remains to lower bound $h(z)/f(z)$:
$$ h(z) = \frac{\int \tilde{\alpha}_{(0)}(u)\varphi(z-u)dG(u)}{\int \tilde{\alpha}_{(0)}(u)dG(u)} \geq \frac{\tilde{c}}{1+\tilde{c}}\cdot \int \varphi(z-u)dG(u) =\frac{\tilde{c}}{1+\tilde{c}} \cdot f(z).$$
Applying the triangle inequality and putting everything together:
\begin{equation*}
\abs{\frac{d}{dz}\mu_{(0)}(z)} \leq \inf_{\tilde{c} > 0 } \cb{ (1+\tilde{c})\cdot\p{\sqrt{-\log(2\pi f^2(z))} +  \sqrt{-\log\p{\frac{2\pi  \tilde{c}^2}{(1+\tilde{c})^2} f^2(z)}}}}.
\end{equation*}
Taking $\tilde{c}=1+\sqrt{2}$ and noting that $2(1+\tilde{c}) < 7$ leads to the bound:
$$\abs{\frac{d}{dz}\mu_{(0)}(z)} \leq 7 \sqrt{-\log(\pi f^2(z))}.$$
Continuing, the second derivative of $\mu_{(0)}(z)$ is equal to:
$$\mu_{(0)}''(z) = \p{\mu_{(0)}(z)+\tilde{c}}\cdot \cb{ \p{\frac{h''(z)}{h(z)}+1} - \p{\frac{f''(z)}{f(z)}+1}} - 2 \mu'_{(0)}(z)\cdot\frac{f'(z)}{f(z)}.$$
Applying Lemma A.1. in~\citet{jiang2009general} a second time we find that:
$$0 \leq \frac{f''(z)}{f(z)} + 1 \leq  -\log(2\pi f^2(z)),\;\; 0 \leq \frac{h''(z)}{h(z)} + 1 \leq  -\log(2\pi h^2(z)).$$
Using the fact that \smash{$\abs{\mu_{(0)}(z)+\tilde{c}} \leq 1+\tilde{c}$}, that we already bounded \smash{$|\mu_{(0)}'(z)|$}, \smash{$f'(z)/f(z)$} above, and the triangle inequality we conclude.\\

\noindent Lower bound: Without loss of generality, we consider the case that $z=0$. Let $\delta_u$ denote the point mass measure at $\cb{u}$. We take $G = \frac{1-w}{2}\cdot\p{\delta_{-t} + \delta_{t}} + w \cdot \delta_0$ for parameters $w \in [0,1]$, $t>0$ which we will specify later and $\alpha_{(0)}(u) = \ind(u=0)$. Then:
$$f(z) = \frac{1-w}{2}\cdot\p{\varphi(z-t) + \varphi(z+t)} + w \cdot \varphi(z),\; \mu_{(0)}(z) =  w \cdot \frac{\varphi(z)}{f(z)}.$$
To simplify notation we write $\mu(\cdot) = \mu_{(0)}(\cdot)$. By direct calculation we can verify that
$$ \mu''(0) = - w \frac{\varphi(0)f(0) + \varphi(0)f''(0)}{f^2(0)},\;\; f''(0) = (1-w)(t^2-1)\varphi(t) - w\varphi(0).$$
Next choose $w = \varphi(t)$, so that $f(0)=  (1+\varphi(0)-\varphi(t))\varphi(t)$ and
$$ \mu''(0) = - \varphi(0) \frac{(1-\varphi(t))\cdot t^2}{(1+\varphi(0)-\varphi(t))^2}.$$
Finally, we pick $t$ so that $\varphi(t) = \rho$, that is, $t=\sqrt{-\log(2\pi \rho^2)}$. It then holds in particular that $f(0) \geq \rho$ and using the fact that $\varphi(t) \in (0, 1/\sqrt{2\pi}]$, we get:
$$ \abs{\mu''(0)} \geq \frac{\varphi(0)(1-\varphi(0))}{(1+\varphi(0))^2} t^2 \geq \frac{1}{10}t^2 =  \frac{1}{10}(-\log(2\pi \rho^2)).$$
\end{proof}

\subsection{Lemma A.1. in~\citet{jiang2009general}}

\setcounter{proposition}{0}
\renewcommand{\theproposition}{A.\arabic{proposition}.}

\begin{lemma}[\citet{jiang2009general}]
Let $G$ be a distribution on $\RR$, let $\varphi$ be the standard normal density function and let $f_{G}(z) = \int \varphi(z-u)dG(u)$ be the density of the convolution $G \star \varphi$. Then:
$$0 \;\leq\; \p{\frac{f_G'(z)}{f_G(z)}}^2 \;\leq\;\frac{f_G''(z)}{f_G(z)} + 1 \;\leq\; -\log\p{2\pi f_G^2(z)}.$$
\end{lemma}
\begin{proof}
Let $U \sim G$ and $Z \mid U \sim \nn(U, 1)$. We may verify the following three equalities:
$$
\begin{aligned}
&\EE{U - z \mid Z=z}=\frac{f_G'(z)}{f_G(z)},\;\; \EE{(U-z)^2 \mid Z=z} = \frac{f_G''(z)}{f_G(z)} + 1,\;\\
&\EE{\sqrt{2\pi}\exp\p{(U-z)^2/2} \mid Z=z} = 1/f_{G}(z).
\end{aligned}
$$
Then, by Jensen's inequality:
$$\p{\frac{f_G'(z)}{f_G(z)}}^2 = \EE{U - z \mid Z=z}^2 \leq \EE{(U-z)^2 \mid Z=z} = \frac{f_G''(z)}{f_G(z)} + 1.$$
Next, define the convex function $h(x) = \sqrt{2\pi}\exp(x/2)$ with inverse $h^{-1}(y) = \log(y^2/(2\pi))$. Applying Jensen's inequality again, we see that:
$$
\frac{f_G''(z)}{f_G(z)} + 1
\leq  h^{-1}\p{\EE{h\p{(U-z)^2} \mid Z=z}} \;=\; h^{-1}(1/f_{G}(z)) 
 = -\log\p{2\pi f_G^2(z)}.
$$
\end{proof}

\section{Balancing the response functions in the absence of effective randomization}
\label{sec:robustness_balancing}

We next prove a robustness guarantee under strong assumptions on $\mu_{(w)}(\cdot)$ when there is no randomization at all---but we nevertheless proceed pretending Assumptions~\ref{assu:noise} and~\ref{assu:exogenous} hold. 
We emphasize, however, that Assumptions~\ref{assu:noise} and~\ref{assu:exogenous} are central to the mechanics and the interpretation of our approach, and so we do not expect our approach to be robust to substantial misspecification of the noise distribution. 

The falsely posited noise model (in Assumption~\ref{assu:noise}) has density ``$p(z \mid u)$'' with respect to ``$\lambda$.''
We assume that $Z_i$ has a density with respect to $\lambda$ which is equal to $f$ and that $\mu_{(w)}(z) = \EE{Y_i(w) \mid Z_i=z}$ takes a specific nonparametric functional form as a linear combination of $p(z \mid u)/f(z)$, 
\begin{equation}
  \label{eq:mu_w_model}
\mu_{(w)}(z) = a_0 + \tau_c w + \frac{\int a(u)p(z \mid u)dH(u)}{f(z)},
\end{equation}
for some $a_0 , \tau_c \in \RR$, a distribution $H$ and a function $a(\cdot)$. Under continuity of $z \mapsto \mu_{(w)}(z)$ at $c$, $\tau_c$ is precisely the causal effect at the cutoff defined in~\eqref{eq:continuity}.

Our main result is analogous to Theorem~\ref{theo:consistency} and Corollary~\ref{coro:bias_decomposition}. We derive the asymptotic limit of $\htau_\gamma$ defined in~\eqref{eq:weighted} for fixed $\gamma_{\pm}$ (that does not depend on $n$)\footnote{It is also possible to derive results under triangular array asymptotics where $\gamma_{\pm}=\gamma_{\pm}^{(n)}$ depends on $n$, analogous to Section~\ref{sec:inference}.} and provide a representation for its asymptotic bias in estimating $\tau_c$ in~\eqref{eq:mu_w_model}.

\begin{proposition}
\label{prop:balancing}
Suppose that Assumption~\ref{assu:rdd} holds with $\mathbb E(Y_i^2) < \infty$, that $\gamma_{\pm}(\cdot)$ is bounded, and that $\mathbb E\{\gamma_+(Z_i)\},\, \mathbb E\{\gamma_-(Z_i)\}>0$ (where $\gamma_{\pm}$ is deterministic and does not depend on $n$). 
Also suppose that $Z_i$ has $d\lambda$-density $f(\cdot)$ and that $\mu_{(w)}(z) = \EE{Y(w) \mid Z=z}$ can be represented as in~\eqref{eq:mu_w_model} with bounded $a(\cdot)$.
We erroneously posit Assumption~\ref{assu:noise} with $d\lambda$ noise density ``$p(z \mid u)$'' and Assumption~\ref{assu:exogenous}. 
Then as $n\to \infty$, $\htau_\gamma - \tau_c - b_{\gamma} = o_{\mathbb P}(1)$ with:
$$ 
\begin{aligned}
b_{\gamma} =  \int a(u) \cb{    \frac{\hkernel{\gamma_+}{u}}{\int \gamma_+(z) f(z) d\lambda(z)}  - \frac{  \hkernel{\gamma_-}{u}}{\int \gamma_-(z) f(z) d\lambda(z)}} dH(u).
\end{aligned}
$$
Above, $\hkernel{\gamma_{\pm}}{u}$ is defined as in~\eqref{eq:hkernel_def} with the misspecified noise density ``$p(z \mid u)$.''
\end{proposition}
Before proving this proposition, we provide some interpretation. Suppose we choose $\gamma_{\pm}$ 
such that $\int \gamma_+(z) f(z) d\lambda(z) \approx \int \gamma_-(z) f(z) d\lambda(z) \approx 1$ (as we seek to achieve via constraint~\eqref{eq:integrate_to_1} in optimization problem~\eqref{eq:qp}) and such that the notional ``$U_i$'' are balanced via $\hkernel{\gamma_+}{u} \approx  \hkernel{\gamma_-}{u}$ (as in constraint~\eqref{eq:simpleQP_bias}; also see the paragraph after Proposition~\ref{prop:balance_response}). Then, even though ``$U_i$'' is only notional (that is, Assumptions~\ref{assu:noise}--\ref{assu:exogenous} do not hold in the data generating process), if~\eqref{eq:mu_w_model} holds, we will have that $b_{\gamma} \approx 0$ and so $\htau_\gamma \approx \tau_c$ for large $n$. 

A special case is as follows: suppose $\mu_{(w)}(z)$ is constant as a function of $z$, then representation~\eqref{eq:mu_w_model} holds with $\alpha(\cdot) \equiv 0$. Then $b_{\gamma}$ in Proposition~\ref{prop:balancing} is equal to $0$, $\htau_{\gamma}$ is consistent for the causal effect and our bias assessment is conservative (no matter what noise model we posit).

\begin{proof}[of Proposition~\ref{prop:balancing}]
The convergence in probability follows as in the proof of Theorem~\ref{theo:consistency}, that is, $\htau_{\gamma} - \taugamma = o_{\mathbb P}(1)$,  
where $\taugamma = \mu_{\gamma,+} - \mu_{\gamma,-}$. Without Assumptions~\ref{assu:noise} and~\ref{assu:exogenous}, $\mu_{\gamma,+}, \mu_{\gamma,-}$ are no longer equal to the expressions in~\eqref{eq:tau_gamma} and we consider an alternative decomposition of the asymptotic bias that does not require these assumptions and instead builds on~\eqref{eq:mu_w_model}. We have that:
  
  $$
  \begin{aligned}
  \mu_{\gamma, +} &= \frac{\EE{\gamma_+(Z)Y_i(1)}}{\EE{\gamma_+(Z)}} \\ 
  &= \frac{\EE{\gamma_+(Z)\mu_{(1)}(Z)}}{\EE{\gamma_+(Z)}} \\ 
  &= \alpha_0 + \tau_c + \frac{ \int \gamma_+(z) \int a(u)p(z \mid u)dH(u) d\lambda(z)}{\int \gamma_+(z) f(z) d\lambda(z)} \\
  &= \alpha_0 + \tau_c + \frac{ \int a(u) \int \gamma_+(z)  p(z \mid u)d\lambda(z) dH(u) }{\int \gamma_+(z) f(z) d\lambda(z)} \\ 
  &= \alpha_0 + \tau_c + \frac{ \int a(u) \hkernel{\gamma_+}{u} dH(u) }{\int \gamma_+(z) f(z) d\lambda(z)}
  \end{aligned}
  $$
  In the last step, it is important to emphasize that we apply the definition of $\hkernel{\gamma_+}{u}$ from~\eqref{eq:hkernel_def} with the misspecified noise model ``$p(z \mid u)$.''
  Arguing analogously for $\mu_{\gamma, -}$, we find that

  $$ 
  \begin{aligned}
  \theta_{\gamma} - \tau_c &= \frac{ \int a(u) \hkernel{\gamma_+}{u} dH(u) }{\int \gamma_+(z) f(z) d\lambda(z)} - \frac{ \int a(u) \hkernel{\gamma_-}{u} dH(u) }{\int \gamma_-(z) f(z) d\lambda(z)},
  \end{aligned}
  $$
as claimed.
\end{proof}

\section{Computational details for worst-case bias}
\label{subsec:bias_computation}

\subsection{Notation}
\label{subsec:bias_comp_notation}
In this section we explain how to compute the worst-case bias in \eqref{eq:biasfractional}. The main idea behind our optimization algorithm is to define $A_{(0)}$, resp. $T$ as the signed measure that is absolutely continuous with respect to $G$ with density $\alpha_{(0)}(u)$, resp. $\tau(u)$. We will parameterize the optimization problem in terms of optimization variables that represent $G$, $A_{(0)}$ and $T$. To simplify notation, we define the following linear functionals:
$$
\begin{aligned}
L_{h,+}(H) &= \int \hkernel{u}{\gamma_+} dH(u),\\
L_{h,-}(H) &= \int \hkernel{u}{\gamma_-} dH(u),\\
L_{w}(H) &= \int w(u) dH(u).
\end{aligned}
$$
Then we can write:
\begin{equation}
\label{eq:bias_as_sum_of_ratios}
\operatorname{Bias}\{{\gamma_{\pm},\, \tau_w; \, \alpha_{(0)}(\cdot),\, \tau(\cdot),\,G}\}  = \frac{L_{h,+}(A_{(0)}) + L_{h,+}(T)}{L_{h,+}(G)} - \frac{L_{h,-}(A_{(0)})}{L_{h,-}(G)} - \frac{L_w(T)}{L_w(G)},
\end{equation}
a sum-of-ratios of linear functionals. We propose solving:
\begin{subequations}
\label{eq:fracopt}
\begin{align}
\sup_{G, A_{(0)}, T} \quad  & \frac{L_{h,+}(A_{(0)}) + L_{h,+}(T)}{L_{h,+}(G)} - \frac{L_{h,-}(A_{(0)})}{L_{h,-}(G)} - \frac{L_w(T)}{L_w(G)}
 \\
\textrm{s.t.} \quad & G \in \mathcal{G}_n,    \\
&  0 \leq \frac{dA_{(0)}}{dG}(u) \leq 1 \text{ for all } u, \label{eq:A_0_constraint} \\
&  0 \leq \frac{dT}{dG}(u) \leq 2M \text{ for all } u \label{eq:T_constraint}.
\end{align}
\end{subequations}
We explain why this problem is equivalent to the problem we care about solving in~\eqref{eq:biasfractional}. There are two observations:
\begin{enumerate}
    \item We first show that it suffices to reduce attention to $T(\cdot)$ satisfying~\eqref{eq:T_constraint} instead of more general $T(\cdot)$ that satisfy the heterogeneity constraint in~\eqref{eq:sensi}. Fix $G$, $A_{(0)}$ and $T$ that are feasible for \eqref{eq:biasfractional}. Let $\bar{\tau}$ be such that $\abs{dT(u)/dG - \bar{\tau}} \leq M$ and define \smash{$\widetilde{T} = (M-\bar{\tau}) G + T$}. Then \smash{$d\widetilde{T}(u)/dG = dT(u)/dG + M - \bar{\tau} \in [0,\,2M]$}. Hence \smash{$G, A_{(0)}, \widetilde{T}$} are also feasible. Furthermore, we may check that $(G, A_{(0)}, T)$ and \smash{$(G, A_{(0)}, \widetilde{T})$} lead to the same value of the objective $\operatorname{Bias}\{\gamma_{\pm},\, \tau_w; \, \cdot, \cdot, \cdot\}$ in~\eqref{eq:bias_as_sum_of_ratios}. 
    \item We next show that we may ignore the absolute value in \eqref{eq:biasfractional}. Fix feasible $G$, \smash{$A_{(0)}$} and $T$. Suppose we replace $A_{(0)}$ and $T$ by \smash{$\widetilde{A}_{(0)} = G - A_{(0)}$} and \smash{$\widetilde{T} = 2M G - T$}. Then \smash{$d\widetilde{A}_{(0)}(u)/dG = 1 - dA_{(0)}(u)/dG \in [0,1]$}, and so the constraint~\eqref{eq:A_0_constraint} will continue to be satisfied and similarly, \smash{$d\widetilde{T}(u)/dG = 2M - dT(u)/dG \in [0, 2M]$}, and so the constraint~\eqref{eq:T_constraint} will also continue to be satisfied. Hence $G$, $\widetilde{A}_{(0)}$ and $\widetilde{T}$ also are feasible and we may further check that the objective value switches sign compared to its original value and retains its absolute value. Thus optimization problem~\eqref{eq:fracopt} is implicitly maximizing the absolute value of the objective.
\end{enumerate}

\subsection{Optimizing the sum-of-ratios objective}
We now explain how~\eqref{eq:fracopt} may be solved numerically.\footnote{Such sum-of-ratios optimization problems have been studied in the optimization literature, see e.g., \citet{benson2007simplicial, konno1999minimization} and references therein.} First, we may reduce the number of ratios in the objective by the~\citet{charnes1962programming} transformation, as follows:
\begin{subequations}
\label{eq:fracopt_cc}
\begin{align}
\sup_{\check{G}, \check{A}_{(0)}, \check{T}, \xi} \quad  & L_{h,+}(\check{A}_{(0)}) + L_{h,+}(\check{T}) - \frac{L_{h,-}(\check{A}_{(0)})}{L_{h,-}(\check{G})} - \frac{L_w(\check{T})}{L_w(\check{G})}
 \\
\textrm{s.t.} \quad & \check{G} \in \cb{ \xi \cdot \tilde{G} : \tilde{G} \in \mathcal{G}_n},  \label{eq:gcal_cc}  \\
&  0 \leq \frac{d\check{A}_{(0)}}{d\check{G}}(u) \leq 1 \text{ for all } u, \label{eq:A_0_constraint_cc} \\
&  0 \leq \frac{d\check{T}}{d\check{G}}(u) \leq 2M \text{ for all } u, \label{eq:T_constraint_cc}\\
& L_{h,+}(\check{G}) = 1, \label{eq:unity_cc}\\
& \xi \geq 0. \label{eq:nonneg_cc}
\end{align}
\end{subequations}
The optimization variables are $\xi \geq 0$, $\check{G}$, $\check{A}_{(0)}$, and $\check{T}$. Their interpretation is as follows: $\xi = 1 / L_{h,+}(G)$, $\check{G} = \xi \cdot G$, $\check{A}_{(0)} = \xi \cdot A_{(0)}$, and $\check{T} = \xi \cdot T$. Next consider solving~\eqref{eq:fracopt_cc} subject to the additional linear constraints that
$$L_{h,-}(\check{G}) = \zeta, \;\;L_w(\check{G}) = \kappa,$$ 
for fixed values of $\zeta, \kappa$. In more detail, let:
\begin{subequations}
\label{eq:profiling_subproblem}
\begin{align}
\mathcal{L}(\zeta, \kappa) = \sup_{\check{G}, \check{A}_{(0)}, \check{T}, \xi} \quad  & L_{h,+}(\check{A}_{(0)}) + L_{h,+}(\check{T}) - \frac{1}{\zeta} L_{h,-}(\check{A}_{(0)}) - \frac{1}{\kappa} L_w(\check{T})
 \\
\textrm{s.t. } & \eqref{eq:gcal_cc},\eqref{eq:A_0_constraint_cc}, \eqref{eq:T_constraint_cc},\eqref{eq:unity_cc},\eqref{eq:nonneg_cc}, \\
& L_{h,-}(\check{G}) = \zeta, \\
& L_w(\check{G}) = \kappa.
\end{align}
\end{subequations}
For fixed values of $\zeta,\kappa$, the above is a linear program. Thus we may solve~\eqref{eq:fracopt_cc} by profiling over $\zeta$ and $\kappa$ and repeatedly solving~\eqref{eq:profiling_subproblem}. Formally, let:
\begin{equation}
\label{eq:zeta_range}
\begin{aligned}
&\ubar{\zeta} = \inf_{\check{G}, \check{A}_{(0)}, \check{T}, \xi}  \quad & \cb{ L_{h,-}(\check{G})} \;\; \textrm{s.t. } & \eqref{eq:gcal_cc},\eqref{eq:A_0_constraint_cc}, \eqref{eq:T_constraint_cc},\eqref{eq:unity_cc},\eqref{eq:nonneg_cc},\\
\text{ and  }\; &\bar{\zeta} = \sup_{\check{G}, \check{A}_{(0)}, \check{T}, \xi}  \quad & \cb{ L_{h,-}(\check{G})} \;\; \textrm{s.t. } & \eqref{eq:gcal_cc},\eqref{eq:A_0_constraint_cc}, \eqref{eq:T_constraint_cc},\eqref{eq:unity_cc},\eqref{eq:nonneg_cc}.
\end{aligned}
\end{equation}
Furthermore, for $\zeta \in [\ubar{\zeta}, \bar{\zeta}]$, let:
\begin{equation}
\label{eq:kappa_range}
\begin{aligned}
&\ubar{\kappa}(\zeta) = \inf_{\check{G}, \check{A}_{(0)}, \check{T}, \xi}  \quad & \cb{ L_w(\check{G})} \;\; \textrm{s.t. } & L_{h,-}(\check{G})=\zeta,\; \eqref{eq:gcal_cc},\eqref{eq:A_0_constraint_cc}, \eqref{eq:T_constraint_cc},\eqref{eq:unity_cc},\eqref{eq:nonneg_cc},\\
\text{ and  }\; &\bar{\kappa}(\zeta) = \sup_{\check{G}, \check{A}_{(0)}, \check{T}, \xi}  \quad & \cb{ L_w(\check{G})} \;\; \textrm{s.t. } & L_{h,-}(\check{G})=\zeta,\;\eqref{eq:gcal_cc},\eqref{eq:A_0_constraint_cc}, \eqref{eq:T_constraint_cc},\eqref{eq:unity_cc},\eqref{eq:nonneg_cc}.
\end{aligned}
\end{equation}
Then the worst-case bias we are interested in is equal to:
\begin{equation}
\label{eq:profiling}
\sup\Big\{\sup\cb{ \mathcal{L}(\zeta, \kappa)\;:\;\kappa \in \sqb{\,\ubar{\kappa}(\zeta),\; \bar{\kappa}(\zeta)}}\;:\;\zeta \in \sqb{\,\ubar{\zeta},\;\bar{\zeta}}\Big\}.
\end{equation}

\subsection{Discretization considerations}
\label{subsubsec:discretization}
To turn the above construction into a practical computational algorithm, we need to solve optimization problems~\eqref{eq:profiling_subproblem}, \eqref{eq:zeta_range}, and \eqref{eq:kappa_range}, as well as solve the profiling task~\eqref{eq:profiling}. We will achieve this by finely discretizing. We refer to~\citet[Supplement D]{ignatiadis2019bias} for a more detailed discussion regarding discretization considerations and describe our implementation choices here.

Instead of optimizing over the space of all distributions for the latent variable $U$, we optimize over all distributions supported on the equidistant finite grid from $a_{\text{min}}$ to $a_{\text{max}}$ with $B+1$ points: 
\begin{equation}
\label{eq:finite_grid}
\mathcal{K}(B, a_{\text{min}}, a_{\text{max}}) = \cb{ a_{\text{min}},\,a_{\text{min}} + \frac{a_{\text{max}}-a_{\text{min}}}{B},\, a_{\text{min}} + 2\frac{a_{\text{max}}-a_{\text{min}}}{B},\,\dotsc,\, a_{\text{max}}}.
\end{equation}
Our default choice uses $B=499$, $a_{\text{min}} = \min\cb{Z_1,\dotsc,Z_n}-2\nu$, $a_{\text{max}} = \max\cb{Z_1,\dotsc,Z_n}+2\nu$ for Gaussian noise distributions with variance $\nu^2$ and $B=399$, $a_{\text{min}} = 10^{-4}$ and $a_{\text{max}} = 1-10^{-4}$ for Binomial noise (the latter choice of $a_{\text{min}}$, $a_{\text{max}}$ for the Binomial empirical Bayes problem is used by default in~\citet{koenker2017rebayes}).

By enumerating the grid elements as $\mathcal{K}(B, a_{\text{min}}, a_{\text{max}}) = \cb{u_1, \dotsc, u_{B+1}}$, we may represent every distribution $G$ supported on this set by the probabilities \smash{$g_j = \PP[G]{U = u_j}$} assigned to $u_j$. The $g_j$ lie on the probability simplex. Furthermore, we may represent $\check{G}$ by $\check{g}_j$, which satisfy:
$$ \sum_{j=1}^{B+1} \check{g}_j = \xi, \; \check{g}_j \geq 0.$$
Analogously, we may represent $\check{T}, \check{A}_{(0)}$ by $(B+1)$-dimensional vectors and we only apply the constraints in~\eqref{eq:A_0_constraint_cc} and ~\eqref{eq:T_constraint_cc} for $u \in \mathcal{K}(B, L, U)$. Hence, after the aforementioned discretization, all of ~\eqref{eq:profiling_subproblem}, \eqref{eq:zeta_range}, and \eqref{eq:kappa_range} turn into finite-dimensional linear programs that we optimize using the MOSEK solver~\citep{mosek}.

To solve the profiling problem~\eqref{eq:profiling}, instead of considering all $\zeta \in [\ubar{\zeta}, \bar{\zeta}]$, we only consider $\zeta \in \mathcal{K}(49, \ubar{\zeta}, \bar{\zeta})$. Meanwhile, for each such $\zeta$ we discretize $[\ubar{\kappa}(\zeta), \bar{\kappa}(\zeta)]$ as an equidistant grid with distance between grid points of at most $\ubar{\zeta}/5$. Hence we solve the discretized~\eqref{eq:profiling} by solving a finite number of linear programs.

\section{Worst-case curvature}

\subsection{Computational details}
\label{subsec:curvature_comp}

The construction for optimizing the worst-case curvature is very similar to the construction in Supplement~\ref{subsec:bias_computation}, i.e., after profiling we reduce the optimization problem to a sequence of linear programming tasks. We provide a sketch here.

Our starting point is the ratio representation of $\mu_{(w)}(\cdot)$,
\begin{equation}
\label{eq:noise_conv}
\mu_{(w)}(z) = \frac{\int \alpha_{(w)}(u)p(z\mid u)\,dG(u)}{\int p(z\mid u)\,dG(u)}.
\end{equation}
We omit the subscript ``${(w)}$'' henceforth and write $\mu(\cdot) = \mu_{(w)}(\cdot)$. We may write
$$ \mu(z) = \frac{N(z)}{D(z)},$$
where $N(\cdot)$, resp. $D(\cdot)$ are the numerator, resp. denominator in~\eqref{eq:noise_conv}. Then, assuming $N(\cdot), D(\cdot)$ are twice differentiable, we get by the chain rule that:
$$ \mu''(z) = \frac{N''(z)}{D(z)} - 2 \frac{N'(z)D'(z)}{D^2(z)} - \frac{N(z)D''(z)}{D^2(z)} + 2 \frac{N(z)(D'(z))^2}{D^3(z)}.$$
The absolute value of the above is the quantity we seek to maximize. The critical observation now is that all of $D(z), D'(z), D''(z)$ are linear functionals of $G$. Similarly, $N(z), N'(z), N''(z)$ are linear functionals of $A$, defined as the measure $\ll G$ with $dA(u)/dG= \alpha_{(w)}(u)$. Applying the~\citet{charnes1962programming} transformation (as we did in Supplement~\ref{subsec:bias_computation}) we may rescale the optimization variables $G$ and $A$ as $\check{G}, \check{A}$, such that $\int p(z \mid u) d\check{G}(u) = 1$. Writing $\check{N}(\cdot), \check{D}(\cdot)$ for the corresponding numerator and denominator
$$\check{N}(\cdot) = \int p(\cdot \mid u) d\check{A}(u),\;\; \check{D}(\cdot) = \int p(\cdot \mid u) d\check{G}(u),$$
we get by the above transformation that $\check{D}(z)=1$, and hence:
\begin{equation}
\label{eq:chain_rule}
\mu''(z) = \check{N}''(z) - 2  \check{N}'(z)\check{D}'(z) - \check{N}(z)\check{D}''(z)  + 2\check{N}(z)(\check{D}'(z))^2.
\end{equation}
To conclude we use a profiling argument as in Supplement~\ref{subsec:bias_computation}. Concretely, fix $\kappa, \zeta$ and consider the linear (in the optimization variables) constraints:
$$ \check{D}'(z) = \zeta,\; \check{D}''(z) = \kappa.$$
Under these constraints, we have that:
$$ \mu''(z) = \check{N}''(z) - 2 \zeta \check{N}'(z) - \kappa \check{N}(z)  + 2\zeta^2 \check{N}(z).$$
This objective is linear in the optimization variables (for fixed values of $\zeta,\kappa$) and so we can maximize/minimize it with respect to the constraints by linear programming.

\subsection{Interpretation and computation of worst-case curvature for binomial running variable}
\label{subsec:curvature_binomial}
In our simulations of Section~\ref{subsec:discrete} we require a curvature upper bound for $B$ in order to apply \texttt{optrdd}.
More concretely, we seek a curvature upper bound analogous to~\eqref{eq:worst_case_curvature} that is justified by the noise model, which in this case follows the binomial distribution ($Z_i \mid U_i \sim \mathrm{Binomial}(K, U_i)$). Note that $p(z\cond u)$ and $\mu_{(w)}(z)$
are only defined at $z \in \cb{0,\dotsc,K}$ and so $\mu''(c)$ and $\Curv(c,\,f_{G}(c),\,p)$ in~\eqref{eq:worst_case_curvature} are ill-defined. 
However, as explained by \citet{kolesar2018inference} and \citet{imbens2019optimized}, 
inference using optrdd with bound $B$ is valid as long as there exists any function interpolating 
$\mu_{(w)}(\cdot)$ at $z \in \cb{0,\dotsc,K}$ that is twice differentiable with worst-case curvature bounded by $B$.
In our computation of $\Curv(c,\,f_{G}(c),\,p)$ we interpolate $p(z \cond u)$ for $z \in (0, K)$ (and consequently $\mu_{(w)}(z)$ through~\eqref{eq:noise_conv}) as $p(z \cond u) = p_B(u; z + 1, \,  K - z + 1) - \log(K+1)$, where $p_B(u; \alpha, \beta)$ is the density of the $\text{Beta}(\alpha, \beta)$ distribution at $u \in (0,1)$.

\end{document}